\documentclass[printer]{aa}

\bibpunct{(}{)}{;}{a}{}{,}

\usepackage{graphicx}
\usepackage{txfonts}
\usepackage{soul}
\usepackage{txfonts}
\usepackage{color}
\usepackage{ulem}

\newcommand{\hii}{H\textsc{ii}}
\def\ks{km s$^{-1}$}
\def\d{$^\circ$}

\def\s{$^{\prime\prime}$}

\def\cm3{cm$^{-3}$}

\def\2{$^{12}$CO}
\def\3{$^{13}$CO}
\def\8{C$^{18}$O}
\def\msol{M$_\odot$}

\def\cm2{cm$^{-2}$}

\begin{document}

%\title{A detailed study of a MYSO jet and the molecular environment using near-IR integral field spectroscopy and millimeter data}

\title{Studying a precessing jet of a massive young stellar object within a chemically rich region}

\author {S. Paron \inst{1}
\and D. Mast \inst{2}
\and C. Fariña \inst{3,4}
\and M. B. Areal \inst{1}
\and M. E. Ortega \inst{1}
\and N. C. Martinez \inst{1}
\and M. Celis Peña \inst{5,6} 
}

\institute{CONICET - Universidad de Buenos Aires. Instituto de Astronom\'{\i}a y F\'{\i}sica del Espacio,
              Ciudad Universitaria, Buenos Aires, Argentina
             \email{sparon@iafe.uba.ar}
\and Observatorio Astronómico, Universidad Nacional de Córdoba, Laprida 854, X5000BGR Córdoba, Argentina
\and Isaac Newton Group of Telescopes, E38700, La Palma, Spain
\and Instituto de Astrof\'{\i}sica de Canarias (IAC) and Universidad de La Laguna, Dpto. Astrof\'{\i}sica, Spain
\and Instituto de Matem\'atica, F\'isica y Estad\'istica, Universidad de Las Am\'ericas, Avenida Rep\'ublica 71, Santiago, Chile.
\and Facultad de Ingeniería, Universidad del Desarrollo, Avenida la Plaza 680, Santiago, Chile.
}

\offprints{S. Paron}

   \date{Received <date>; Accepted <date>}

\abstract
{}
{In addition to the large surveys and catalogs of massive young stellar objects (MYSOs) and outflows, dedicated studies of
particular sources, in which high-angular observations (mainly at near-IR and
(sub-)mm) are analyzed in deep, are needed to shed light on the processes involved in the formation of massive stars.
The galactic source G079.1272+02.2782, a MYSO at about 1.4 kpc of distance that appears in several catalogs, is an ideal source to carry out this kind of studies (hereafter MYSO G79). This is because of its relatively close distance, and the intriguing structures that the source shows at near-IR wavelengths.} 
{Near-IR integral field spectroscopic observations were carried out using NIFS at Gemini-North. The spectral and angular resolutions, about 2.4--4.0 \AA, and 0\farcs15--0\farcs22, allow us to perform a detailed study of the source and its southern jet, resolving structures with sizes between 200 and 300 au. As a complement, millimeter data retrieved from the James Clerck Maxwell Telescope and the IRAM 30\,m telescope databases were analyzed to study the molecular gas around the MYSO at a larger spatial scale.}
{
The detailed analysis of a jet extending southwards MYSO G79 shows cork-screw like structures at 2.2 $\mu$m continuum, strongly suggesting that the jet is precessing. The jet velocity is estimated between 30 and 43 \ks~and its kinematic indicates that it is blue-shifted, i.e. the jet is coming to us along the line of sight.
We suggest that the precession may be produced by the gravitational tidal effects generated in a probable binary system, and we estimate a jet precession period of about 10$^{3}$ yr, indicating a slow-precessing jet, which is in agreement with the observed helical features.
An exhaustive analysis of H$_{2}$ lines at the near-IR band along the jet allows us to investigate in detail a bow-shock produced by this jet. We find that this bow-shock is indeed generated by a C-type shock and it is observed coming to us, with some inclination angle, along the line of sight. This is confirmed by the analysis of molecular outflows at a larger spatial scale. A brief analysis of
several molecular species at millimeter wavelengths indicates a complex chemistry developing at the external layers of the molecular clump in which MYSO G79 is embedded.
We point out that we are presenting an interesting observational evidence that can give support to theoretical models of bow-shocks and precessing jets. }{}

\titlerunning{Activity of a high-mass YSO}
\authorrunning{S. Paron et al.}


\maketitle
%
%-------------------------------------------------------------------

\section{Introduction}

It is known that massive young stellar objects (MYSOs) are short lived and 
they form deeply embedded in dense molecular clumps. In general, the relevant phases of
massive stars formation are very difficult to be observed, and hence they remain in the darkness preventing our complete knowledge about star formation. However, theoretical and
observational studies support the idea that MYSOs might be born as a scaled-up version of their low-mass counterparts (e.g. \citealt{beu07,tan14,frost19}). 
%Thus, detailed observational studies of MYSOs can shed light on this issue. 

Indeed, as in low-mass YSOs, structures in Keplerian rotation have been discovered in massive protostars and they
were associated with accretion discs or toroids (e.g. \citealt{beltran16}). Additionally, jets related to MYSOs were
also discovered \citep{mcleod18}, and powerful molecular outflows related to cores containing massive protostars were extensively surveyed  (\citealt{davis04,caratti15,maud15b} and references therein).

In order to probe gaseous discs, cavities generated by jets, shocked 
and ionized gas at the close surroundings of massive protostars we need to observe at near-IR bands. For instance, the analysis of the CO overtone
bandheads, H$_{2}$ and Br$\gamma$ emission lines, in addition to the continuum emission at these wavelengths, is very useful to carry out
deep studies of such structures and/or processes related to massive star formation \citep{hoff06,martins10,fedriani20}. The analysis of the H$_{2}$ lines and the ratios 
among them is useful to discern the excitation mechanisms responsible for the emission of this molecule, that can be excited due to collisions produced by shock fronts or fluorescence excitation by non ionizing UV photons in the Lyman-Werner band (912-1108 \AA). These mechanisms can be distinguished since they preferentially
populate different levels producing different line ratios \citep{davis03,martin-h08}. Moreover, using spectroscopy at near-IR bands is possible to trace the jets kinematic, estimate radial velocities and/or study velocities fields. And the near-IR imaging allows us to eventually 
indirectly analyze tidal forces and torques which might be visible as three-dimensional effects in the jet structure and jet propagation. Indeed, there are observational
evidence and theoretical models that indicate nonaxisymmetric features like jet precession or curved ballistic motion of the jet suggesting that the jet source is part
of a binary system or even a multiple system (see \citealt{sheik22}, and references therein). 

Nowadays there are useful surveys of MYSOs, molecular outflows, and extended near-IR H$_{2}$ 
emission related to high-mass young stars (\citealt{lumsden13,navarete15,caratti15,maud15a,maud15b,yang18}). These surveys provide large and homogeneous samples of sources that allow us to select particular objects to perform dedicated observations for deeper and more detailed studies on individual objects.
Indeed, studies of particular sources (see for instance \citealt{gredel06,fedriani19,fedriani20} and \citealt{areal20}), in which the observations (mainly at near-IR and (sub-)mm) are analyzed in depth, are extremely useful to shed light on the formation processes of MYSOs.  

In this work we investigate MYSO G079.1272+02.2782 (hereafter MYSO G79) and its close surroundings using near-IR integral field spectroscopic observations, which allows us to obtain detailed information from both spectra and photometry. Additionally, using millimeter data retrieved from the James Clerck Maxwell Telescope (JCMT) and the IRAM 30\,m telecope databases, we analyze the molecular gas around MYSO G79 at a larger spatial scale with the aim of complementing the near-IR results.

MYSO G79,  related to IRAS 20216+4107, is an interesting object located at a distance of about 1.4 kpc, a quite close distance for a MYSO, that presents NH$_{3}$ maser emission \citep{urqu11}. This source is included in the distance-limited sample of massive molecular outflows studied by \citet{maud15b} through the emission of the CO J=3--2 line, and its outflows have a total mass of about 8 \msol. \citet{navarete15} detected extended H$_{2}$ emission at near-IR 
towards this object. In fact, this source is catalogued as MHO3476 in the UKIRT Widefield Infrared Survey for H$_2$ (UWISH2), a survey of YSO jets 
in the Galactic plane, particularly towards the Galactic region Cygnus-X \citep{makin18}. According to the catalog,  MHO3476 presents only one H$_{2}$ outflow
lobe whose emission is observed as a collection of knots. A single spectrum at near-IR of MYSO G79 reveals only an H$_{2}$ line  (1--0 S(1) at 2.12 $\mu$m) and some upper values for other lines \citep{cooper13}. This source, also catalogued as a high-mass protostellar object (HMPO) \citep{srid2002}, was observed, among 
59 high-mass star-forming regions, using the IRAM 30\,m telescope at bands 86-94 GHz, 217-221 GHz, and 241-245 GHz \citep{gerner14}. Several molecular lines, such as H$^{13}$CO$^{+}$, HN$^{13}$C, HCN, HCO$^{+}$ J=1--0 among many others, were observed towards MYSO G79. These molecules were used by the authors to study the chemical evolution in the early phases of massive star formation among the analyzed sample. 

\begin{figure}[h]
   \centering
   \includegraphics[width=9cm]{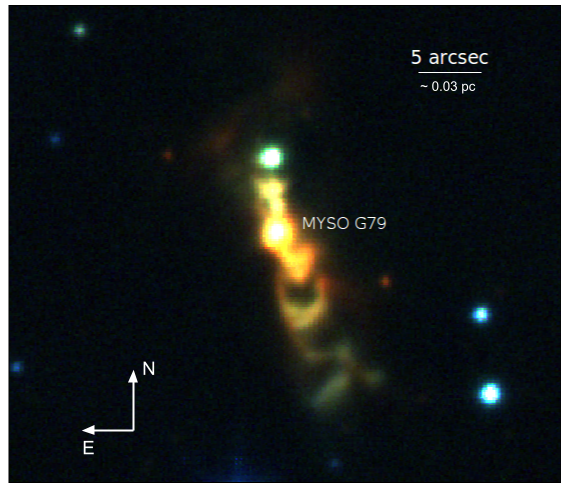}
    \caption{Three-color image with the JHKs emission extracted from the UKIRT Infrared Deep Sky Survey (UKIDSS)
    presented in blue, green, and red respectively. The bar of 5 arcsec represents a size of $\sim0.03$ pc at the assumed
    distance of 1.4 kpc.}
              \label{present}
    \end{figure}

Figure\,\ref{present} displays  a RGB image of the JHKs-band emission obtained from the UKIRT Infrared Deep Sky Survey (UKIDSS). Several arc-like features appears extending from the central object (indicated as MYSO G79 in the figure) to the south, suggesting a cavity carved out by a precessing jet or a helical flow (this will be discussed below). Towards the north some extended near-IR emission also appears and unrelated point-like source,  note its different color, not reddened,
indicating that this source is not so deep embedded like MYSO G79.
Taking into account the relatively close distance, the high galactic latitude which avoids confusion due to the emission and absorption of material along the galactic plane, and the information published in several catalogs and surveys, MYSO G79 is an ideal source to carry out a dedicated study with high angular and spectral resolution and high sensitivity at near-IR using integral field spectroscopic. We focus our study on the central source and the southern extended emission.

 The paper is organized as follows: Sect.\,\ref{dataS} describes the near-IR observations and the data reduction, Sect.\,\ref{resultsS} presents the results obtained from the analysis of the near-IR data, in Sect.\,\ref{discussS} such
results are discussed together with a discussion regarding the molecular gas and chemistry at a larger spatial scale, and finally, in Sect.\,\ref{concl}, concluding remarks are presented.

\section{Observations and data reduction}
\label{dataS}

Near-IR integral field spectroscopic observations were carried out using the Near-infrared Integral Field Spectrograph \citep[NIFS,][]{McGregor2003} located at Gemini North during the first semester of 2019 (project: GN-2019A-Q-108-47). NIFS was used together with the Gemini North Adaptive Optics facility ALTAIR with a natural guide star (NGS). The NGS chosen was TYC 3160-1037-1, a star of unknown spectral type at RA= 20:23:24.12 Dec=+41:17:16.5. With UCAC4 mag V=11.35 (R=11.47) this NGS is well within the V-band limit for full correction regime. ALTAIR was used with field lens IN to improve the Strehl ratios and image quality as the NGS is further than 5 arcsec from all the fields’ centres.

\begin{figure}[h]
   \centering
   \includegraphics[width=8cm]{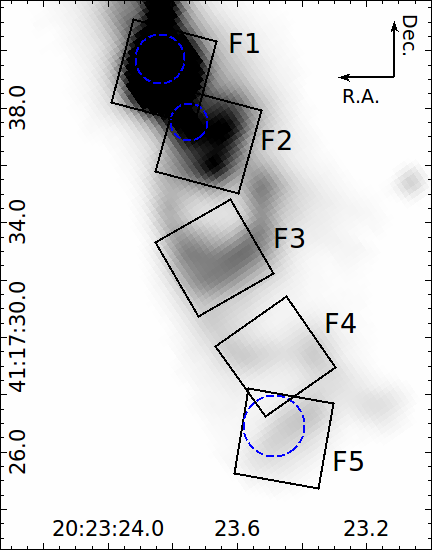}
    \caption{Ks-band emission extracted from the UKIRT Infrared Deep Sky Survey (UKIDSS) towards MYSO G79. Numbered boxes are the fields of 3\s$\times$3\s observed using NIFS at Gemini. Dashed blue circles are the regions from which were extracted the spectra presented below. }
              \label{obs}
    \end{figure}

The K\_G5605 grating (central wavelength: $2.20\,\mu$m, spectral range: $1.99-2.40\,\mu$m, spectral resolution 5290) together with the HK\_G0603 filter (central wavelength: $2.16\,\mu$m, spectral range: $1.57-2.75\,\mu$m) were used to observe five fields that cover the G79 MYSO compact object and
the southern arc-like features. The fields centers and position angles were chosen to enclose the main structures as seeing in the UKIDSS image (see Fig.\,\ref{obs}) in compromise with the separation imposed by the guide star. The position centers of the observed fields, the integration times, the position angle (PA),  and the separation between the NGS and the fields’ centres are presented in Table\,\ref{obsTab}.
 
\begin{table}[h]
\centering
\tiny
\caption{Observed fields using NIFS+ALTAIR at Gemini North.}
 \begin{tabular}{cccccc}
\hline
\hline
Field & RA          & Dec          & Exp. Time  & PA & NGS distance\\
      &             &              &  (sec)    &  (deg) & (arcsec)\\
\hline
1     & 20:23:23.83 & +41:17:39.24 &  25       & 75  &  23.0 \\
2     & 20:23:23.69 & +41:17:36.83 &  85       & 75  &   20.9 \\
3     & 20:23:23.67 & +41:17:32.76 & 240       & 30  &   17.2  \\
4     & 20:23:23.48 & +41:17:29.32 & 300       & 35  &  14.5   \\
5     & 20:23:23.46 & +41:17:26.47 & 240       & 80  &   12.2 \\
\hline
\label{obsTab}
\end{tabular}
\end{table}

The observations followed the Object-Sky dither sequence, with off-source sky positions, and spectra centered at 2.2 $\mu$m. The spectral range is 2.009--2.435 $\mu$m. The spectral resolution ranges from 2.4 to 4.0 \AA, as determined from the full width at half maximum (FWHM) measurement of the ArXe lamp lines used to calibrate the wavelength of the spectra. This results in velocity resolutions in the range 33--55 km s$^{-1}$. The error in the determination of the radial velocities will depend on the signal-to-noise ratio (S/N) of the measured lines (this is described in Sect.\,\ref{spect}). The angular resolution is in the range of 0\farcs15--0\farcs22, derived from the FWHM of the flux distribution of telluric standard stars, which corresponds to 0.0010--0.0015 pc (206--310 au) at the distance of G79.  As there are no point sources in the fields, it is not possible to measure the final correction made by the ALTAIR system. Anyway, the distances reported in Table \ref{obsTab} between the NGS and the fields, and the magnitude of the NGS, both within the expected limits to obtain a close-to-full correction, indicate that the spatial resolution achieved in the fields corresponds to the range mentioned above.

The standard NIFS tasks included in the Gemini IRAF package v1.14\footnote{http://www.gemini.edu/sciops/data/software/gemini\_v114\_rev.txt} were used for data reduction.The procedure included image trimming, flat-fielding, sky subtraction, wavelength calibration, and S-distortion correction.  The telluric correction of the fields was performed by observing two telluric standards: HIP 102074 for Fields 1-4 and HIP 103159 for Field 5. The telluric correction process included fitting the hydrogen absorption lines of the stellar spectrum and fitting a synthetic blackbody spectrum to recover the correct shape of the spectral distribution. These telluric stars were also used to flux calibrate the datacubes. Finally, data cubes were created for each field with a Field-of-View (FoV) of $3\arcsec\times3\arcsec$ and an angular sampling of 0\farcs05$\times$0\farcs05. For the analysis of the datacubes we used the code IFSCUBE\footnote{https://ifscube.readthedocs.io/en/latest/intro.html} \citep{Dutra2021}. For each emission line detected in the datacube, this analysis consists in fitting a continuum to each spaxel and then fitting a gaussian function to the emission line. This way we obtain the flux, the FWHM, the central wavelength, and the continuumm value corresponding to each spaxel in the field. With this information we were able to construct maps of the spatial distribution of these parameters.

\section{Results}
\label{resultsS}

In this section we present the spectra with the line identification, continuum maps, and line emission maps with continuum subtracted of each observed field. To facilitate the interpretation of the maps, the figures are presented with the same inclination (i.e. PA respect to the equatorial coordinates) of the fields shown in Fig.\,\ref{obs}. 

\subsection{Spectra}
\label{spect}

In what follows we present the spectra obtained towards each field in which atomic and molecular lines appear. After a careful inspection of each data cube we find that such lines appear in Fields 1, 2, and 5. Fields 3, and 4 lack of emission lines. The blue dashed circles displayed in Fig.\,\ref{obs} represent the regions in these fields from which the spectra presented below were extracted. These are the regions in which the emission lines are clearly present within the observed field.  

\begin{figure}[h]
   \centering
   \includegraphics[width=9cm]{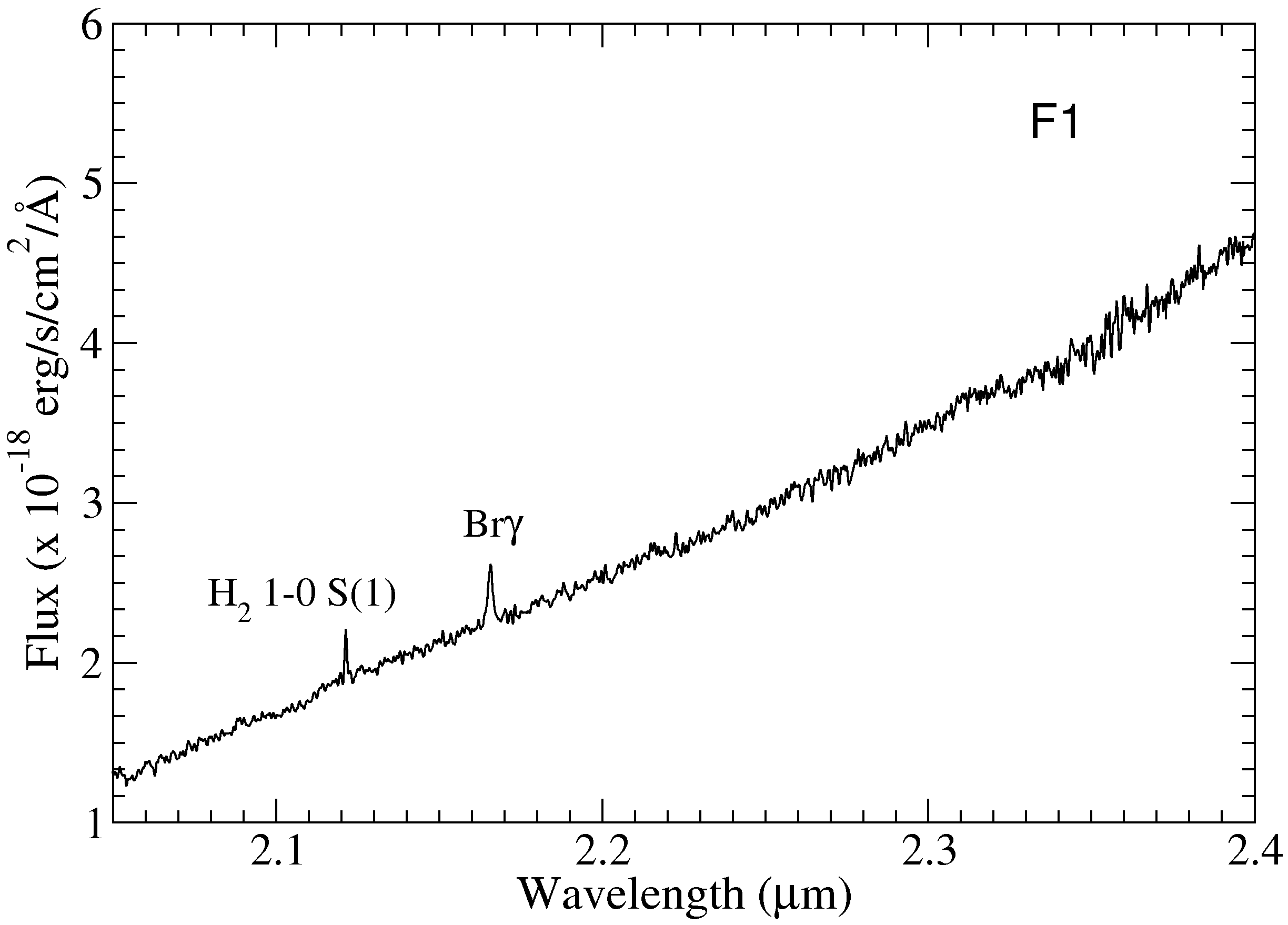}
    \caption{Spectrum obtained towards a region of about 1\farcs6 in diameter (see Fig.\,\ref{obs}) in which the lines appear in Field\,1.
    }
              \label{spectF1}
    \end{figure}

\begin{figure}[h]
   \centering
   \includegraphics[width=9cm]{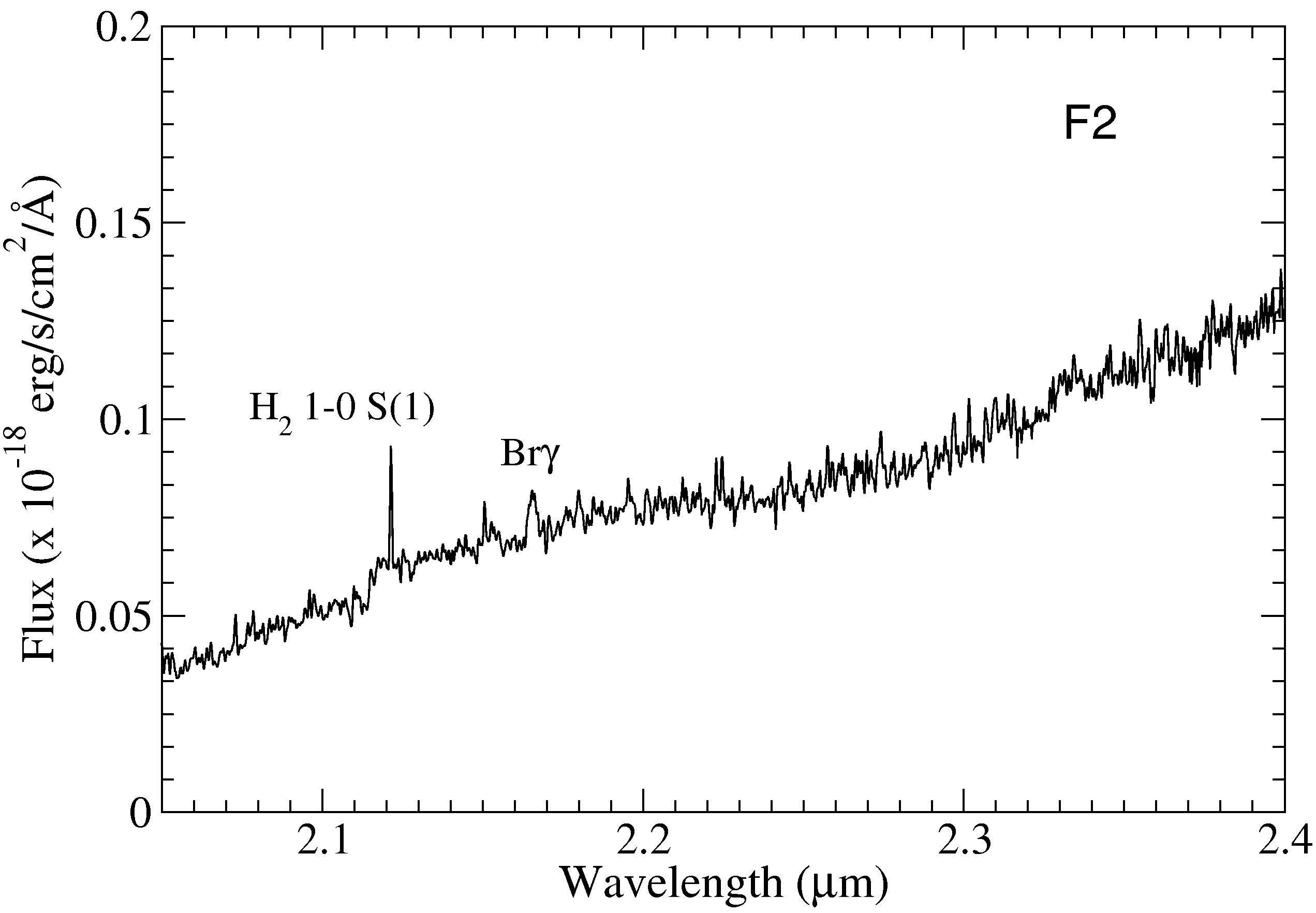}
    \caption{Spectrum obtained from a  region of about 1\farcs2 in diameter towards the upper left corner of Field\,2 (see Fig.\,\ref{obs}).}
              \label{spectF2}
    \end{figure}

\begin{figure}[h]
   \centering
   \includegraphics[width=9cm]{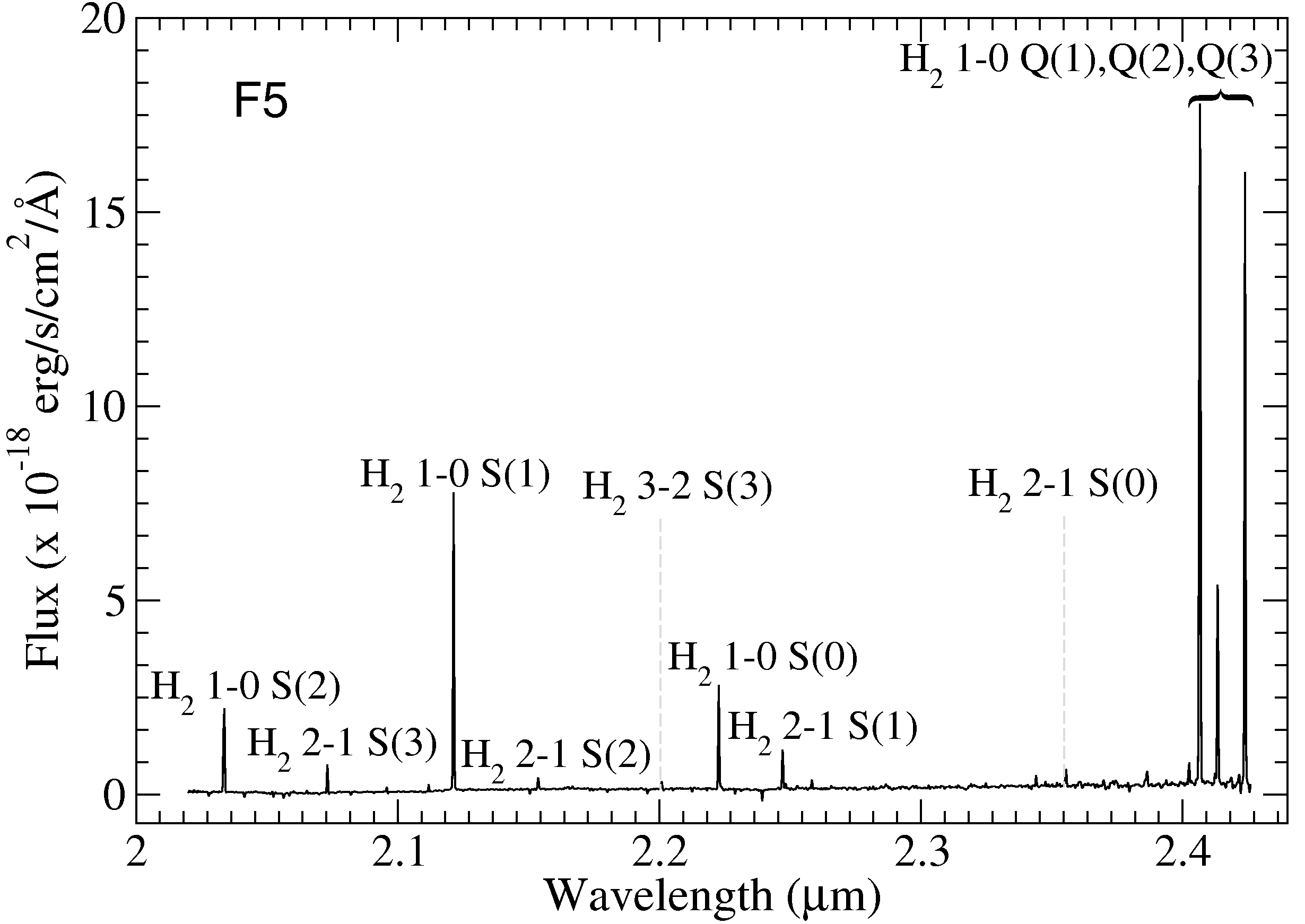}
    \caption{Spectrum obtained from a region of about 2\s~in diameter towards almost the center of Field\,5 (see Fig.\,\ref{obs}).}
              \label{spectF5}
    \end{figure}

Figure\,\ref{spectF1} displays the spectrum in the whole observed wavelength range obtained towards Field\,1 from a circular region (radius about 0\farcs8; see Fig.\,\ref{obs}) at the position in which 
emission lines appear (see the maps Sect.\,\ref{mapsSect}). By inspecting the whole Field\,2 we found only a strong emission line of H$_{2}$ 1--0 S(1) almost at the upper left corner of the observed field (see the circular region of 0\farcs6 in radius in Fig.\,\ref{obs}). 
Also this region presents a Br$\gamma$ marginal emission. Fig.\,\ref{spectF2} shows the spectrum obtained from such a region. Finally, Field\,5 presents very weak continuum emission and it is the richest field among the five observed in H$_{2}$ lines. Fig.\,\ref{spectF5} displays a spectrum obtained towards a region of about 2\s~in diameter lying almost at the center of the field (see Fig.\,\ref{obs}), in which all detected H$_{2}$ lines are indicated.

From the H$_{2}$ 1-0 S(1) emission line in Field 5, we obtain a radial velocity (v$_{\rm LSR}$) for the gas of about $-31$ \ks. This line was chosen for the radial velocity determination for two reasons. It has a S/N of $\sim174$ on the peak emission and $\sim10$ in the lowest part of the map, and also because is near the central wavelength of the spectra which ensures that the wavelength calibration is optimal compared to transitions located at the redder end of the spectrum. This high S/N allows, according to our empiric experience, to obtain radial velocities with a precision of 1/10 the spectral resolution. 
The errors for the reported radial velocities, considering the fitting error and the wavelength calibration, range between $10-15$ \ks. For the estimation of these errors we considered the uncertainty in the continuum fitting and the variations of different lines along the spectra when available.

\subsection{Maps}
\label{mapsSect}

In what follows, we present the continuum and the emission line (continuum subtracted) maps from each field.

\begin{figure}[h]
   \centering
   \includegraphics[width=8.3cm]{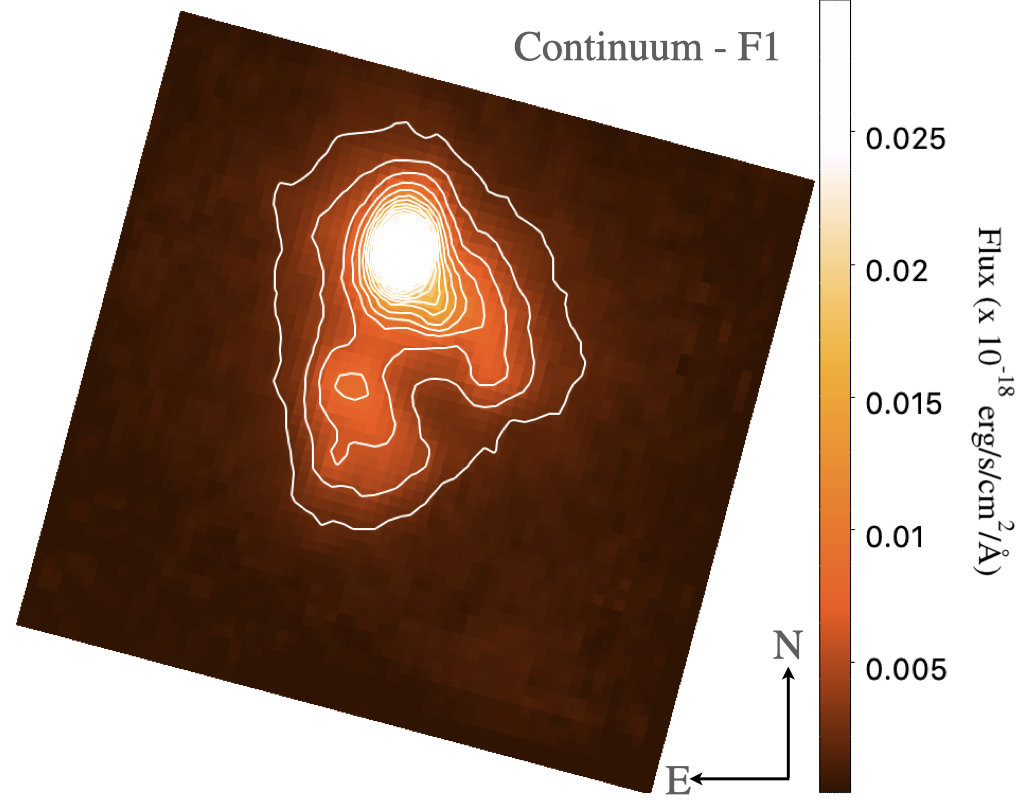}
   \includegraphics[width=8.3cm]{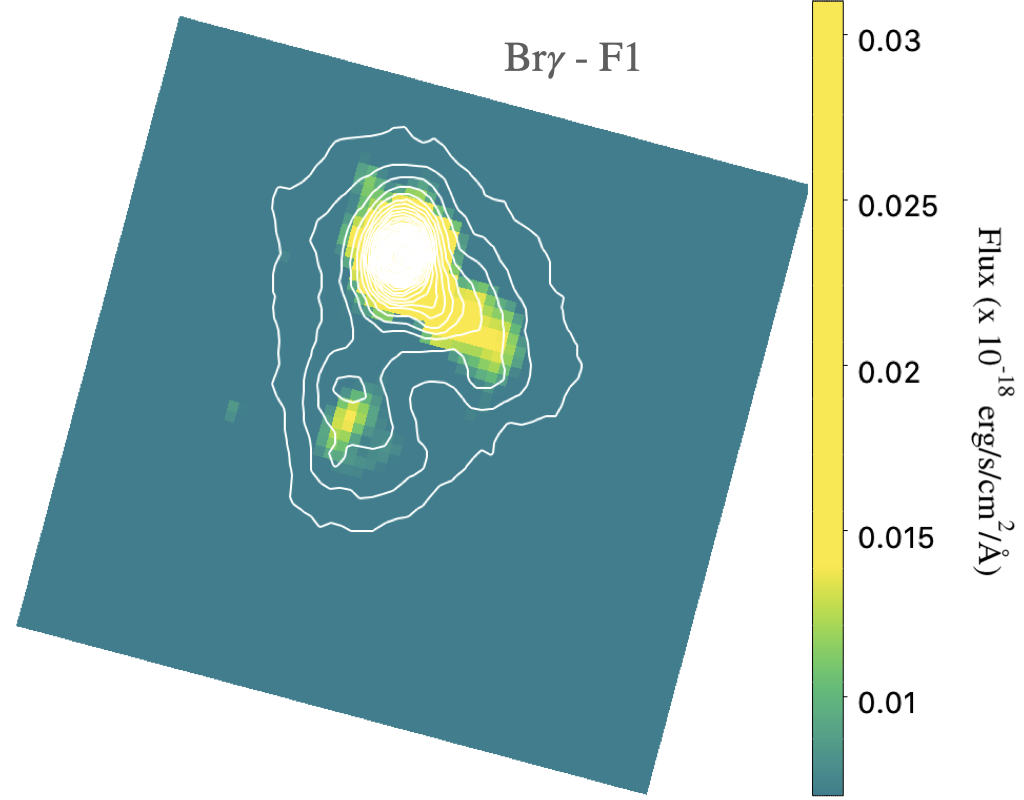}
    \includegraphics[width=8.3cm]{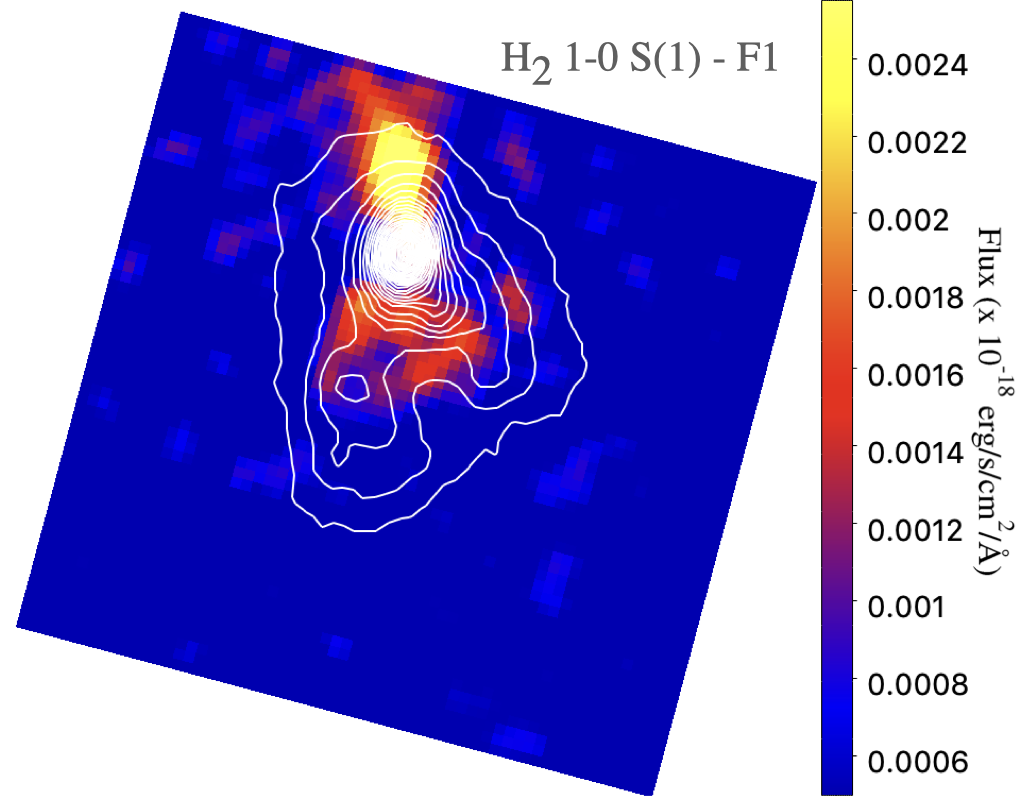}
    \caption{Maps obtained towards Field\,1. The 2.2 $\mu$m continuum map (top) is displayed
    in colour scale with contours that are superimposed for comparison in the Br$\gamma$ (middle) and H$_{2}$ 1--0 S(1) (bottom) line emission maps (continuum subtracted). The background $rms$ is $\sim0.15, 2.97,$ and $0.65\times10^{-23}$ erg/s/cm$^2$/\AA\, for the continuum, the Br$\gamma$, and the H$_2$ maps respectively.}
              \label{mapsF1}
    \end{figure}

\begin{figure}[h]
   \centering
   \includegraphics[width=8.3cm]{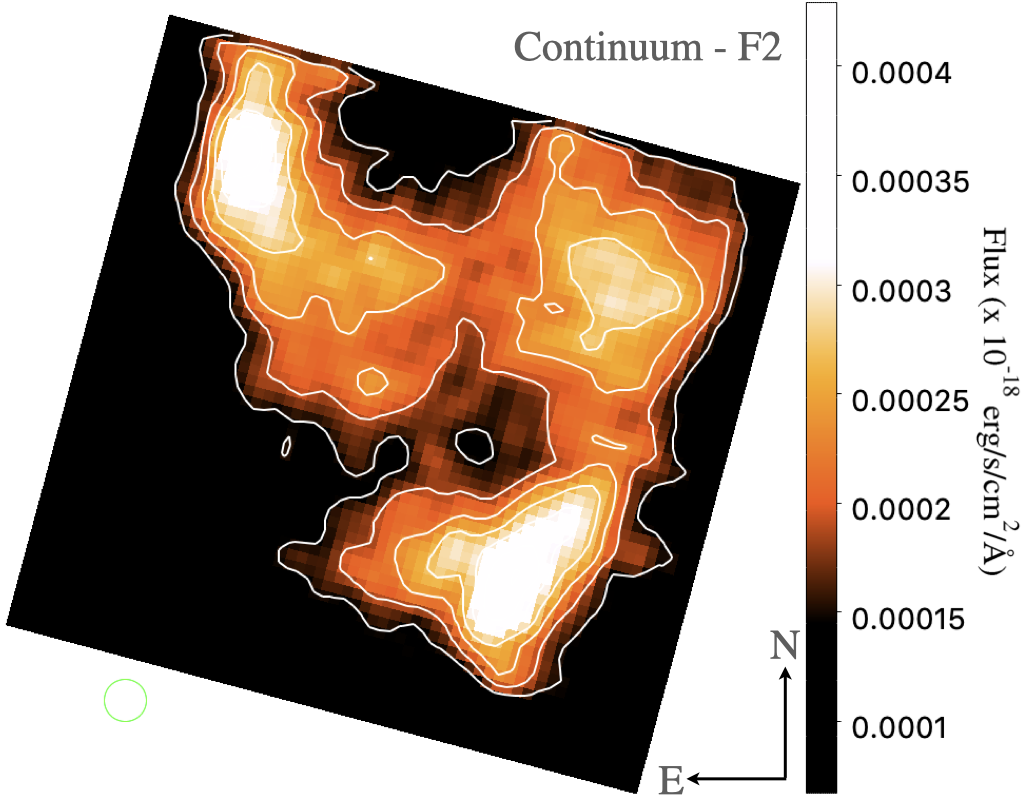}
    \includegraphics[width=8.3cm]{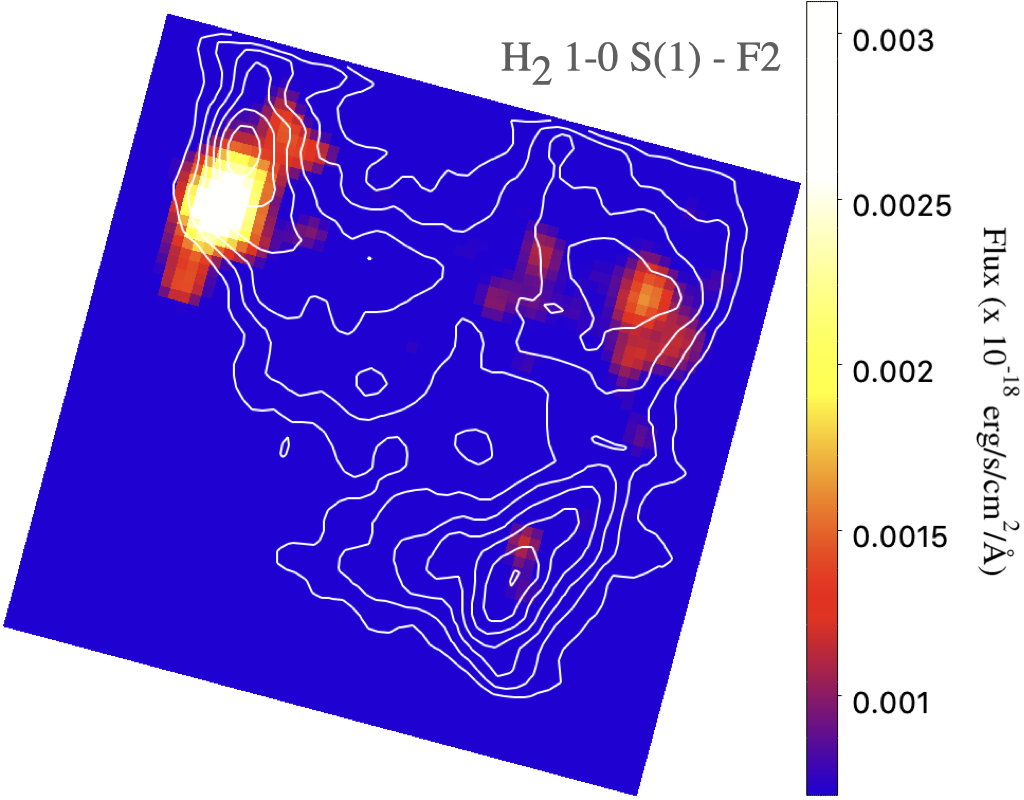}
    \caption{Maps obtained towards Field\,2. The 2.2 $\mu$m continuum map (top) is displayed
    in colour scale with contours that are superimposed for comparison in the continuum subtracted H$_{2}$ 1--0 S(1) line emission map (bottom). The background $rms$ is $\sim2.0$ and $20\times10^{-23}$ erg/s/cm$^2$/\AA for the continuum and the H$_2$ maps respectively.}
              \label{mapsF2}
    \end{figure}

\begin{figure}[h]
   \centering
   \includegraphics[width=8.3cm]{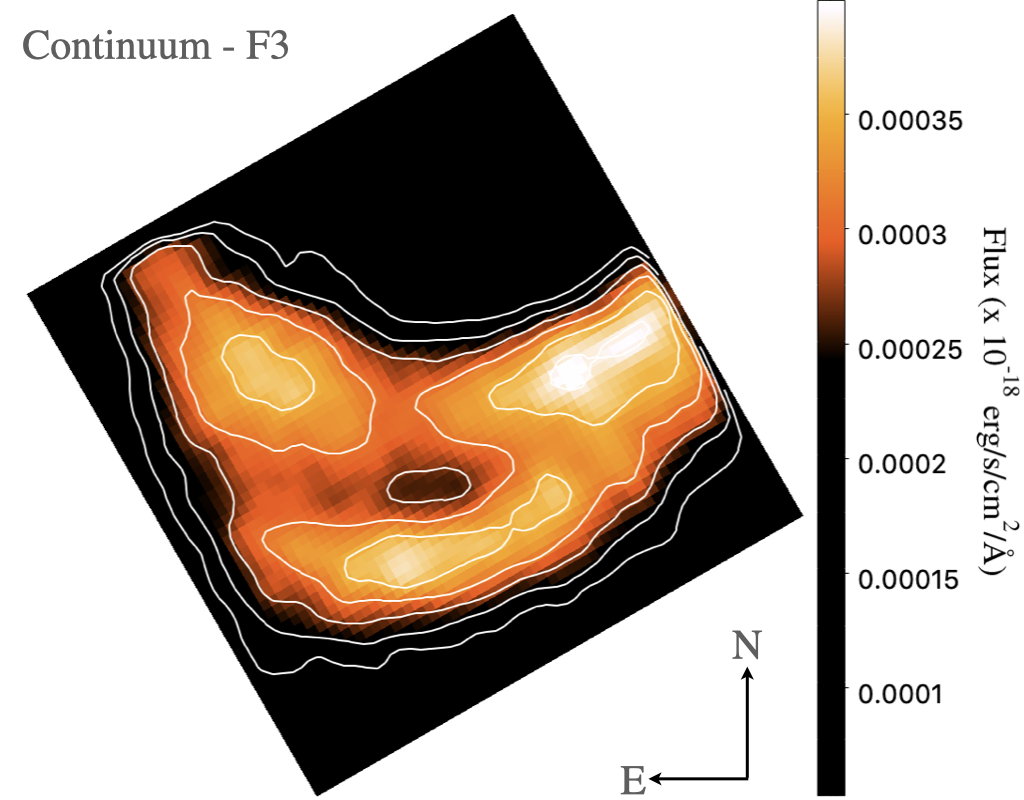}
    \caption{Map of the 2.2 $\mu$m continuum obtained towards Field\,3. The background $rms$ is $\sim0.56\times10^{-23}$ erg/s/cm$^2$/\AA.}
              \label{mapsF3}
    \end{figure}

\begin{figure}[h]
   \centering
   \includegraphics[width=8.3cm]{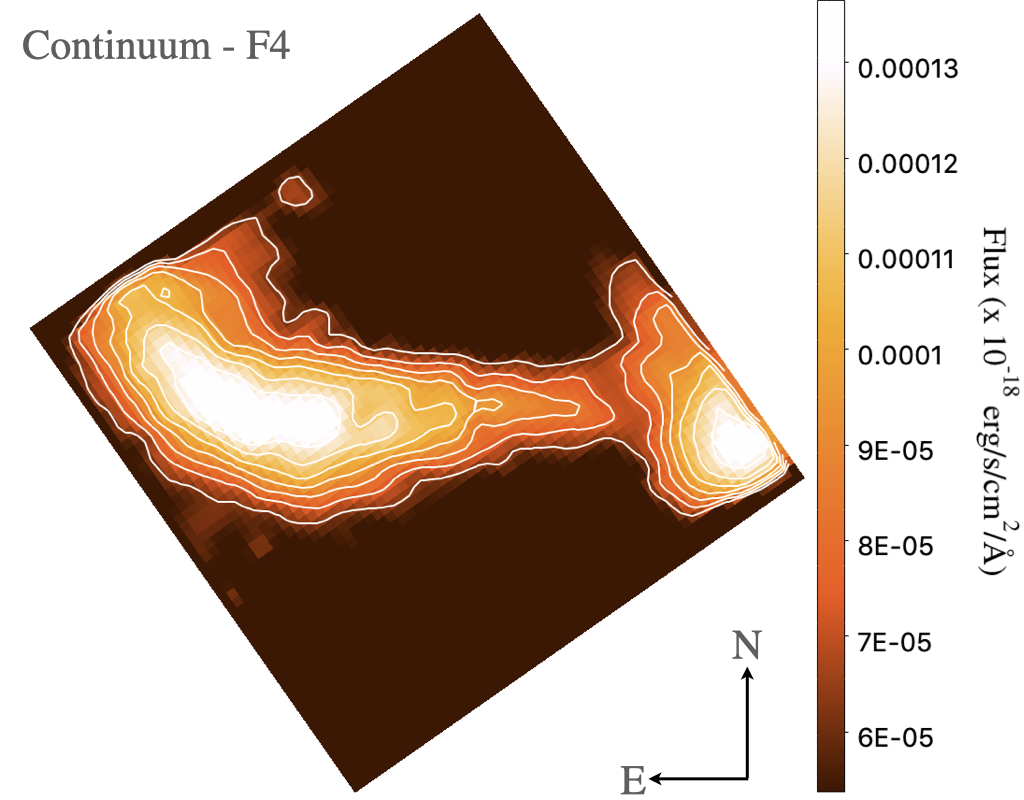}
    \caption{Map of 2.2 $\mu$m continuum obtained towards Field\,4. The background $rms$ is $\sim0.51\times10^{-23}$ erg/s/cm$^2$/\AA.}
              \label{mapsF4}
    \end{figure}

\begin{figure*}[h]
   \centering
   \includegraphics[width=6cm]{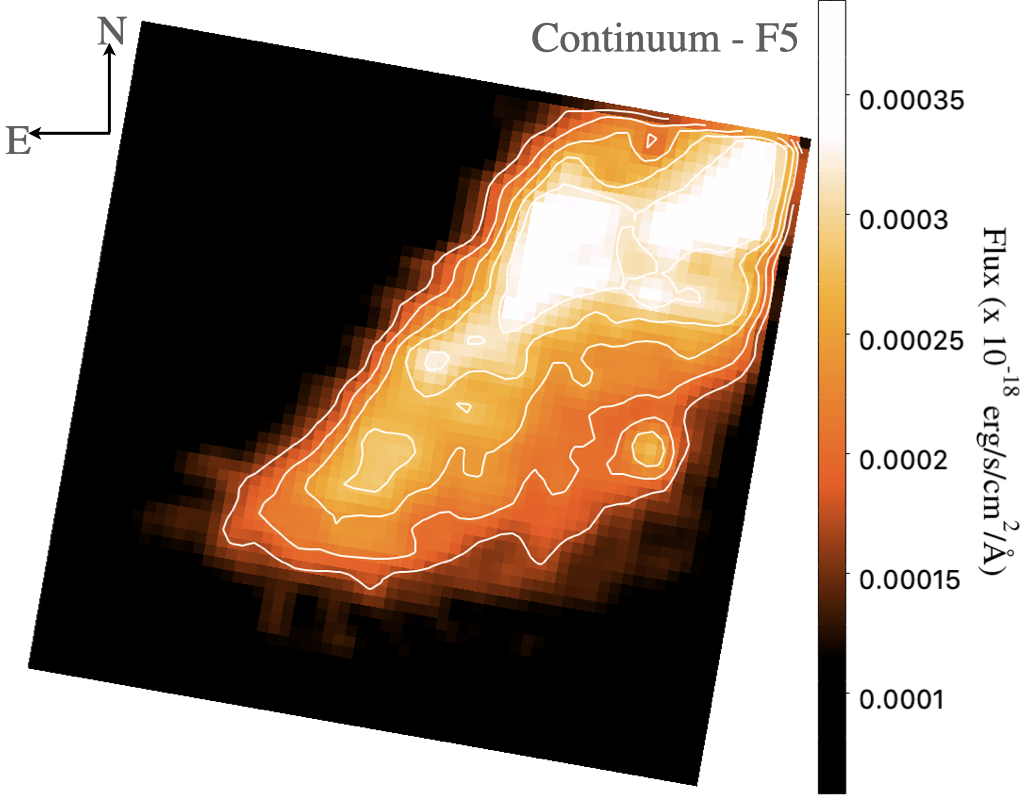}
   \includegraphics[width=6cm]{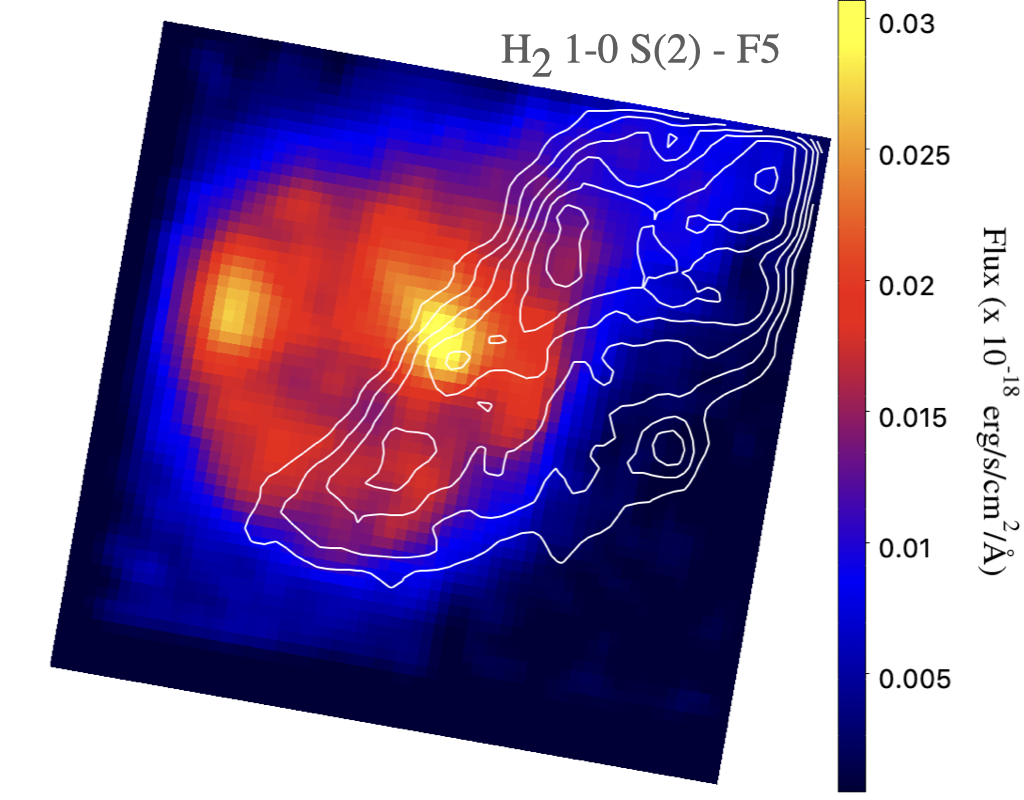}
    \includegraphics[width=6cm]{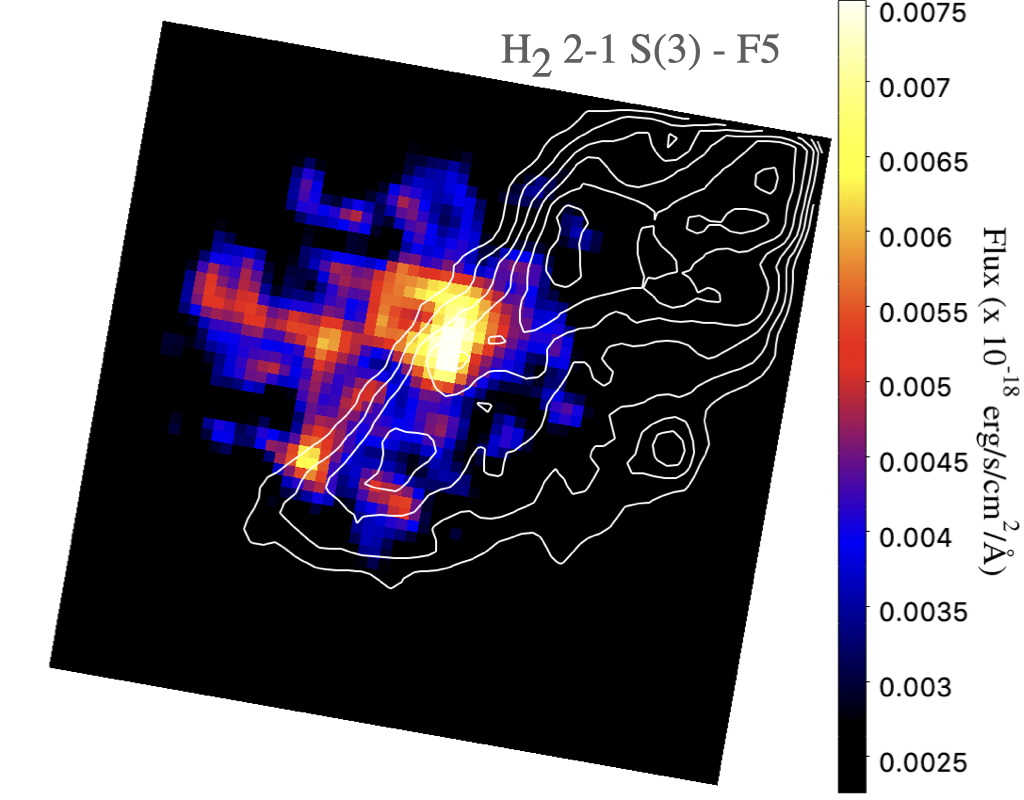}
    \includegraphics[width=6cm]{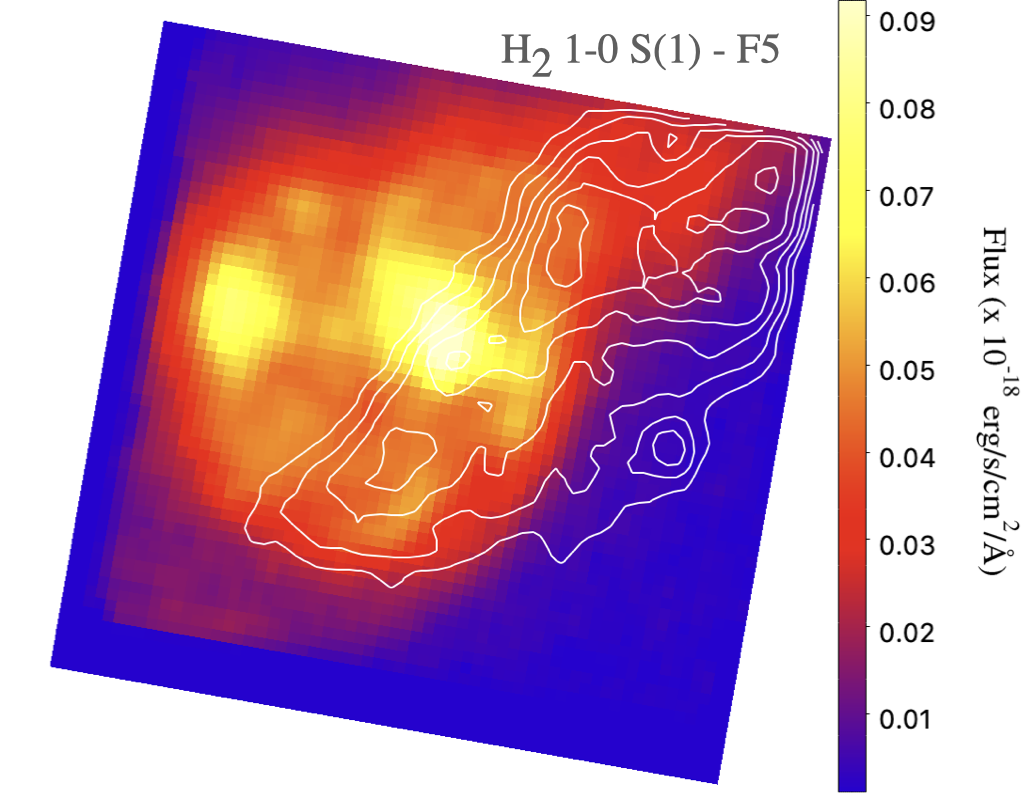}
    \includegraphics[width=6cm]{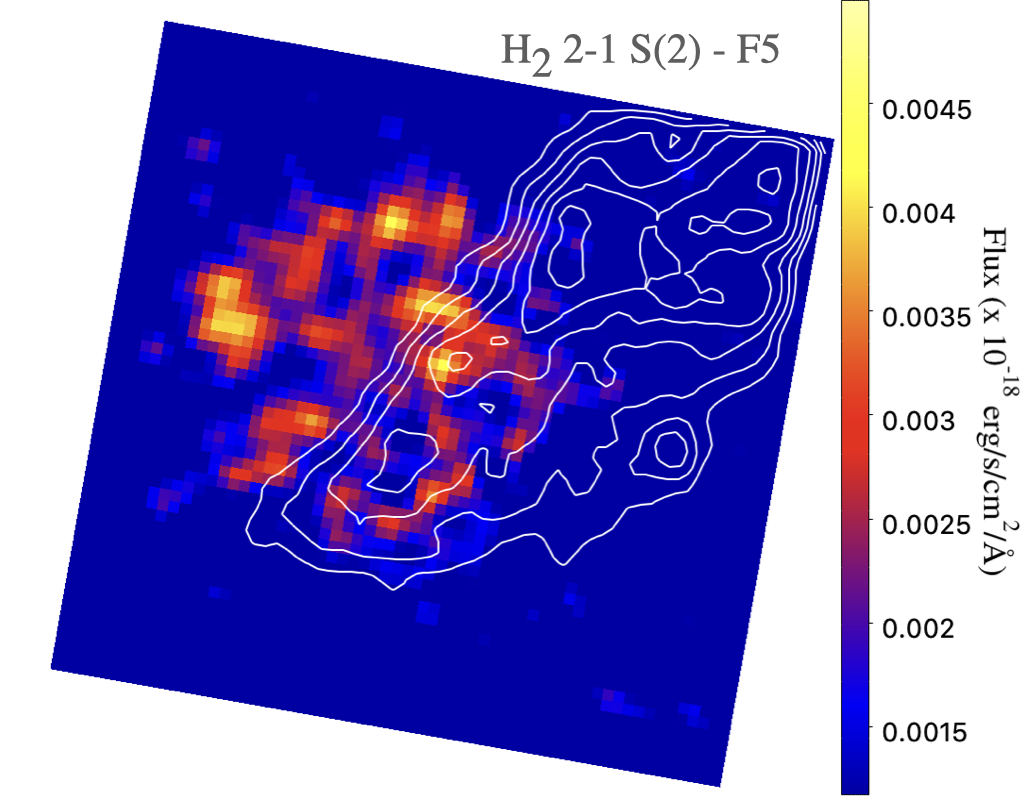}
    \includegraphics[width=6cm]{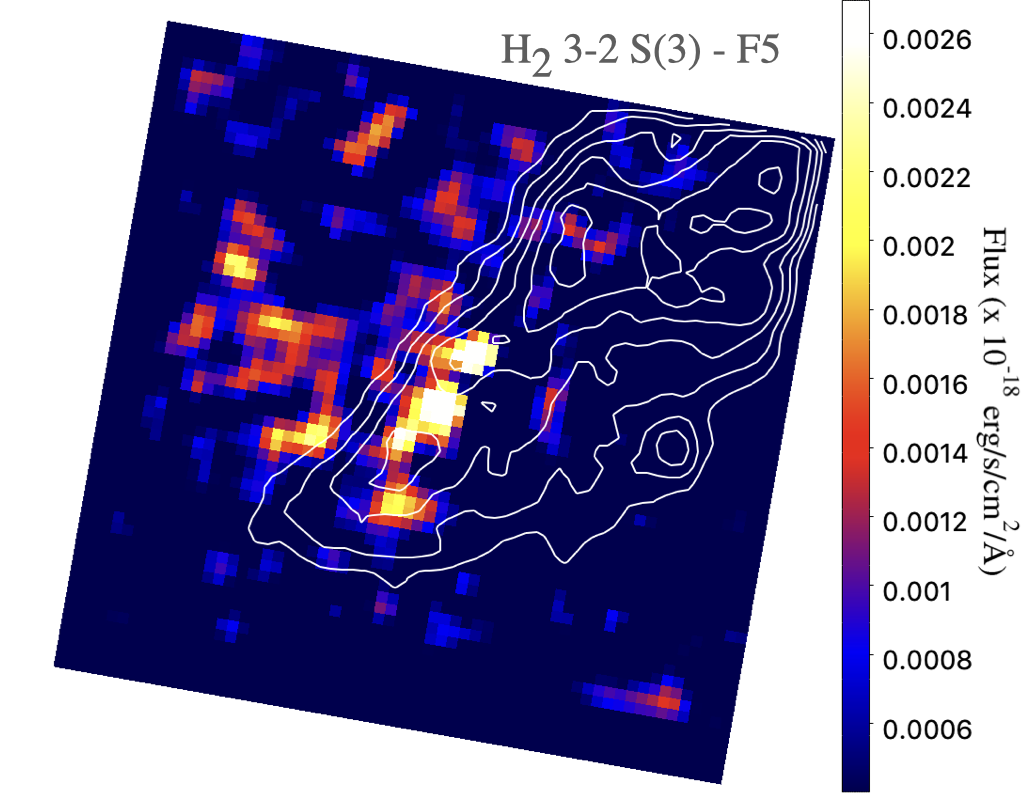}
    \includegraphics[width=6cm]{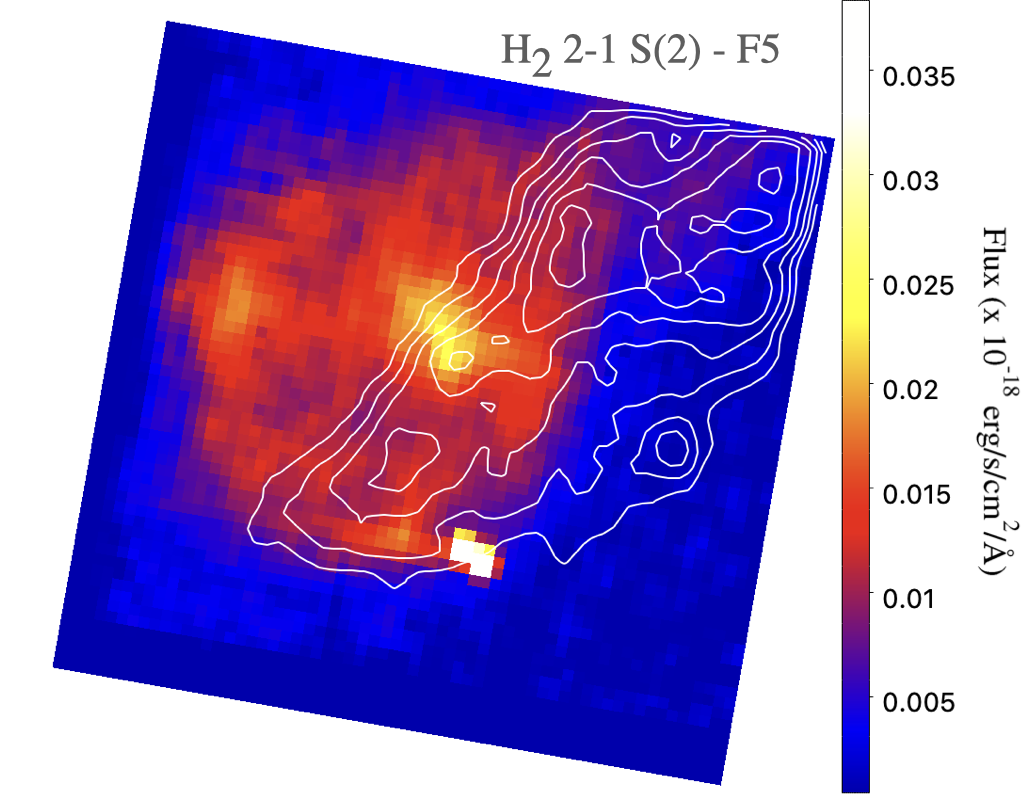}
    \includegraphics[width=6cm]{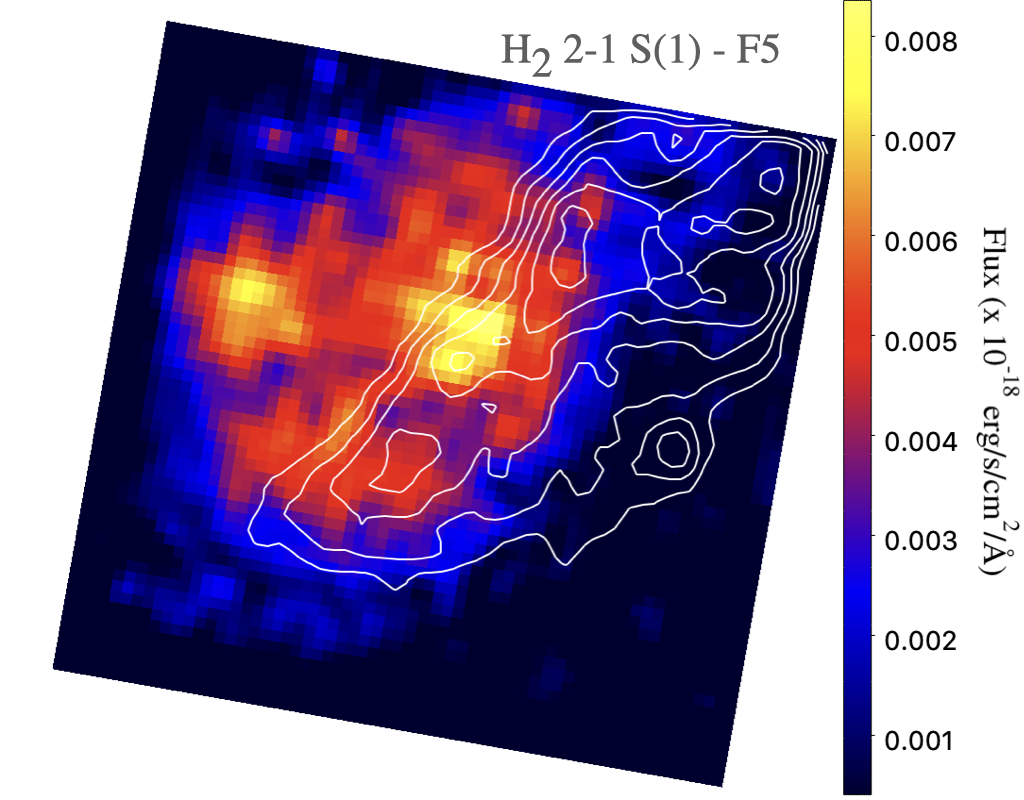}
    \caption{Maps obtained towards Field\,5. The 2.2 $\mu$m continuum map (top-left) is displayed
    in colour scale with contours that are superimposed for comparison in the subsequent continuum subtracted line maps. This figure continues in Fig.\,\ref{mapsF5cont}. The background $rms$ for the corresponding maps are (from top to bottom and left to right) $\sim2.0, 120, 0.92, 0.78, 100, 0.84, 0.93,$ and $0.94\times10^{-23}$ erg/s/cm$^2$/\AA.
    }
              \label{mapsF5}
    \end{figure*}

\begin{figure*}[h]
   \centering
   \includegraphics[width=7cm]{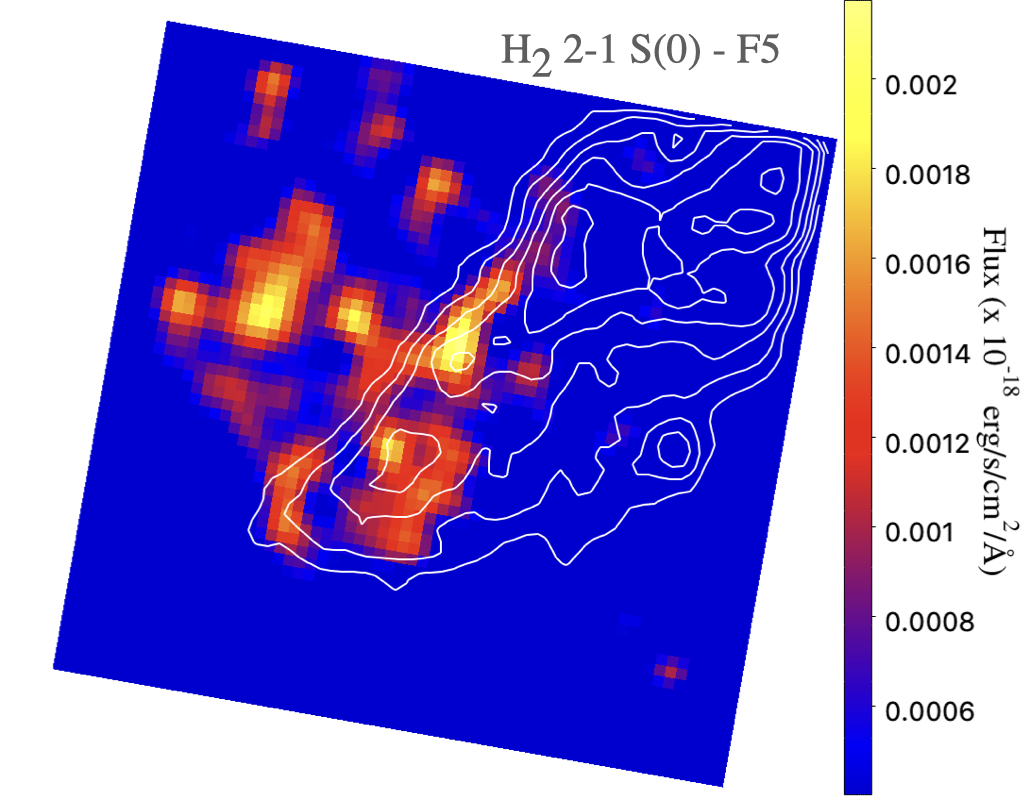}
    \includegraphics[width=7cm]{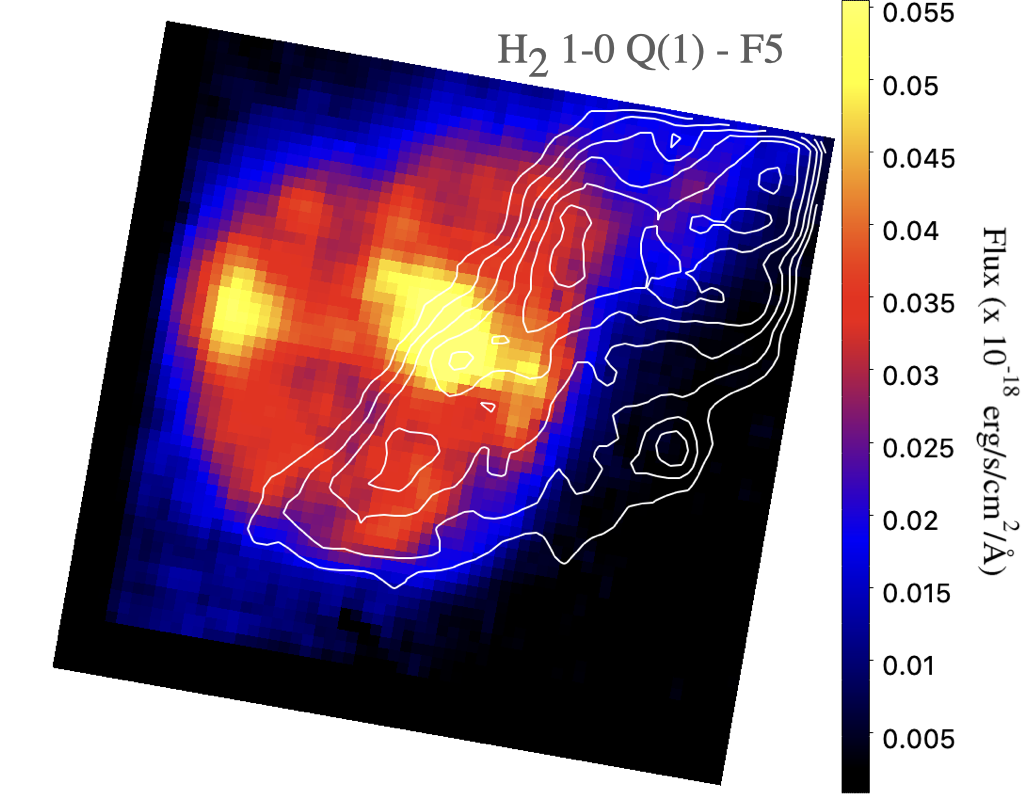}
    \includegraphics[width=7cm]{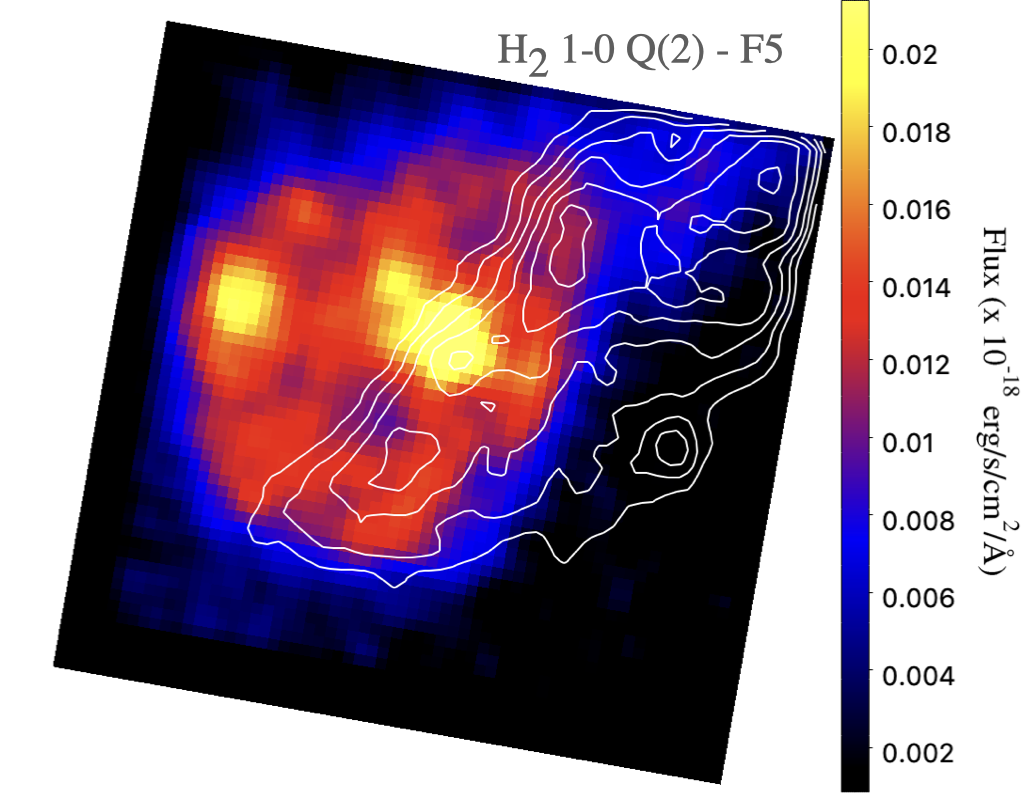}
    \includegraphics[width=7cm]{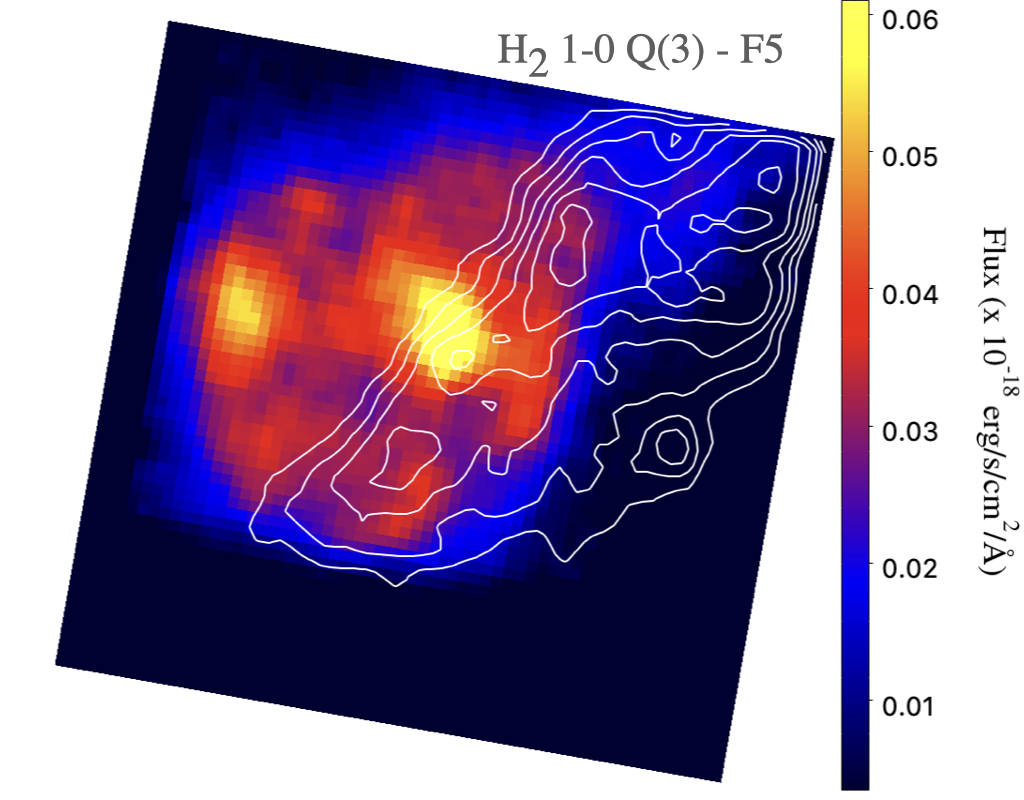}
    \caption{Fig.\,\ref{mapsF5} {\it continued}. The background $rms$ for the corresponding maps are (from top to bottom and left to right) $\sim0.62, 120, 120 $ and $7.9\times10^{-23}$ erg/s/cm$^2$/\AA.
    }
              \label{mapsF5cont}
    \end{figure*}

Fig.\,\ref{mapsF1} displays the 2.2 $\mu$m continuum map,
and the Br$\gamma$ and H$_{2}$ 1--0 S(1) continuum subtracted emission line maps in Field\,1. This field, in which the protostar, understood as the compact object, lies, presents only emission of Br$\gamma$ and H$_{2}$ 1--0 S(1).

Field\,2 possibly marks the beginning of a jet driven by the source located at Field\,1 or a cavity produced by such a jet, that extends towards the south-west (see Fig.\,\ref{present}). The only significant emission line that appears in this field is H$_{2}$ 1--0 S(1). Figure\,\ref{mapsF2} displays the 2.2 $\mu$m continuum map and the mentioned H$_{2}$ line map with the continuum subtracted.

In the case of Fields\,3 and 4, we present only the continuum maps (Figs.\,\ref{mapsF3} and \ref{mapsF4}) because, as mentioned above, no line above the noise level appears in such fields.

\begin{figure}[h]
   \centering
   \includegraphics[width=8.7cm]{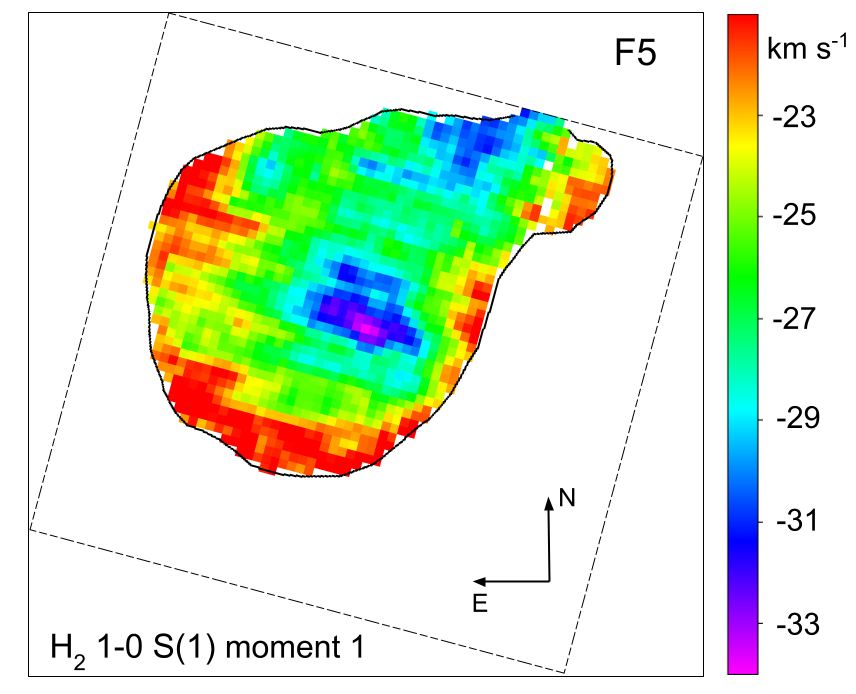}
    \caption{Intensity-weighted mean velocity (moment 1) of the H$_{2}$ 1--0 S(1) line in Field\,5. The central velocity of the line is v$_{\rm LSR} \sim -31$ \ks. A contour (above 4$\sigma$) of the H$_{2}$ emission is displayed just for reference.}
              \label{m1H2F5}
    \end{figure}

Field\,5 likely marks
the end of the jet driven by MYSO G79 or the jet cavity. 
Figures\,\ref{mapsF5} and\,\ref{mapsF5cont} present the 2.2 $\mu$m continuum map and the H$_{2}$ line maps with the continuum subtracted of this field. The molecular emission shows an almost circular morphology uncorrelated with the morphology of the continuum emission. This feature has a radius of about 1\farcs1 ($\sim$1500 au at the distance of 1.4 kpc). Finally, Fig.\,\ref{m1H2F5} displays the intensity-weighted mean velocity (moment 1) of the H$_{2}$ 1--0 S(1) line in Field\,5. As mentioned above, the central velocity of the line is v$_{\rm LSR} \sim -31$ \ks.

\subsection{Line ratios in Fields\,1 and\,5}
\label{ratioSect}

Even though line ratios cannot be used as a sole discriminant of excitation mechanisms of the gas \citep{burton92}, they, together with the 
analysis of the context in which the lines arise, can be used to have a good idea about such mechanisms.

\begin{figure}[h]
   \centering
   \includegraphics[width=8cm]{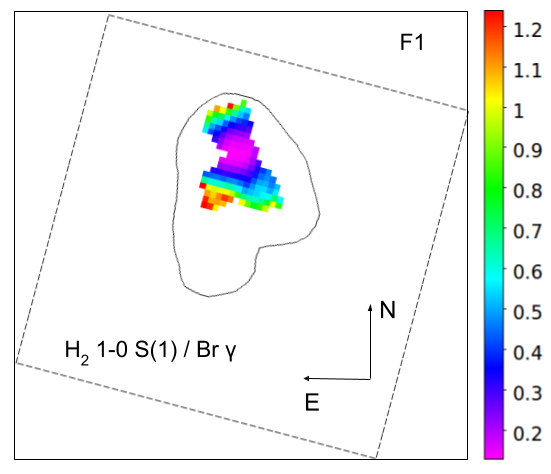}
    \caption{H$_{2}$ S(1) 1--0/Br$\gamma$ ratio in Field\,1. A contour of the continuum emission (above 4$\sigma$) is shown just
    for reference.}
              \label{ratio1}
    \end{figure}

In Field\,1, given that only one H$_{2}$ line appears, we can do just a comparison between the molecular and atomic hydrogen emission through the 
H$_{2}$ 1--0 S(1)/Br$\gamma$ line ratio. This kind of molecular-to-atomic line ratios are commonly used to distinguish between various excitation mechanisms \citep{hatch05,chen15}. Figure\,\ref{ratio1} 
displays a map of such a ratio, in which values above 4$\sigma$ in both lines were used. The average value of the H$_{2}$ 1--0 S(1)/Br$\gamma$ line ratio is 0.6, which will be discussed in Sect.\,\ref{F1}. 

\begin{figure}[h]
   \centering
   \includegraphics[width=7cm]{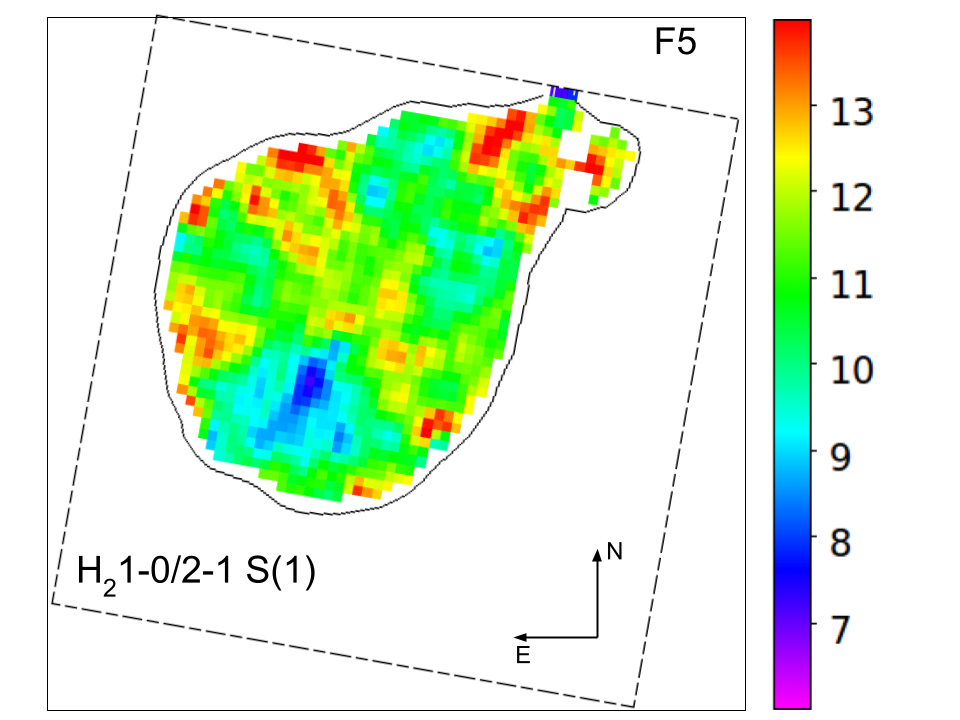}
    \includegraphics[width=7cm]{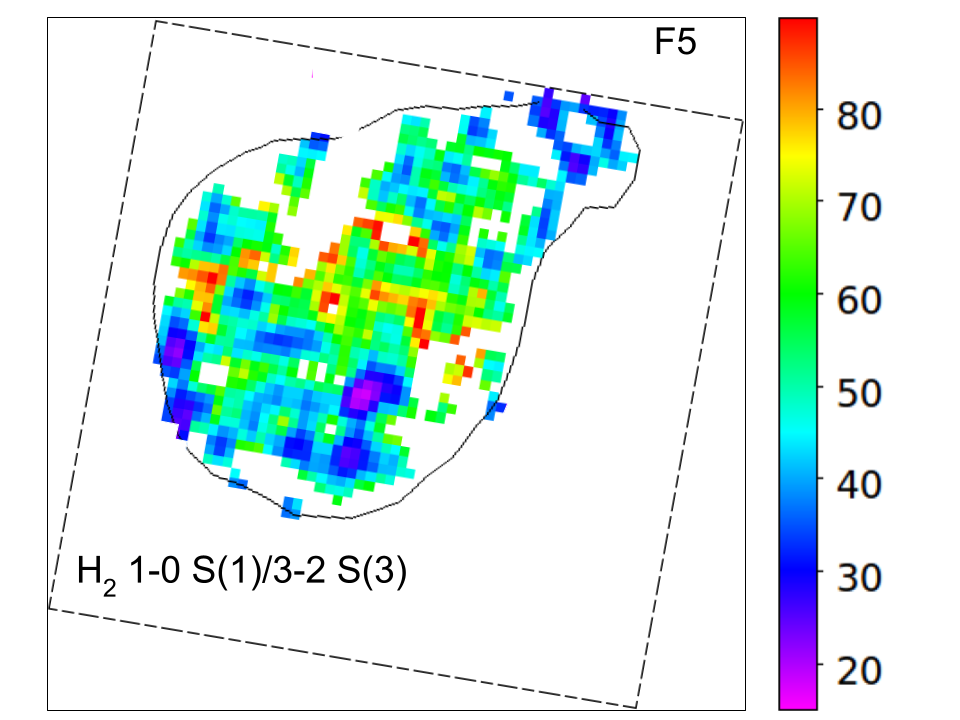}
    \includegraphics[width=7cm]{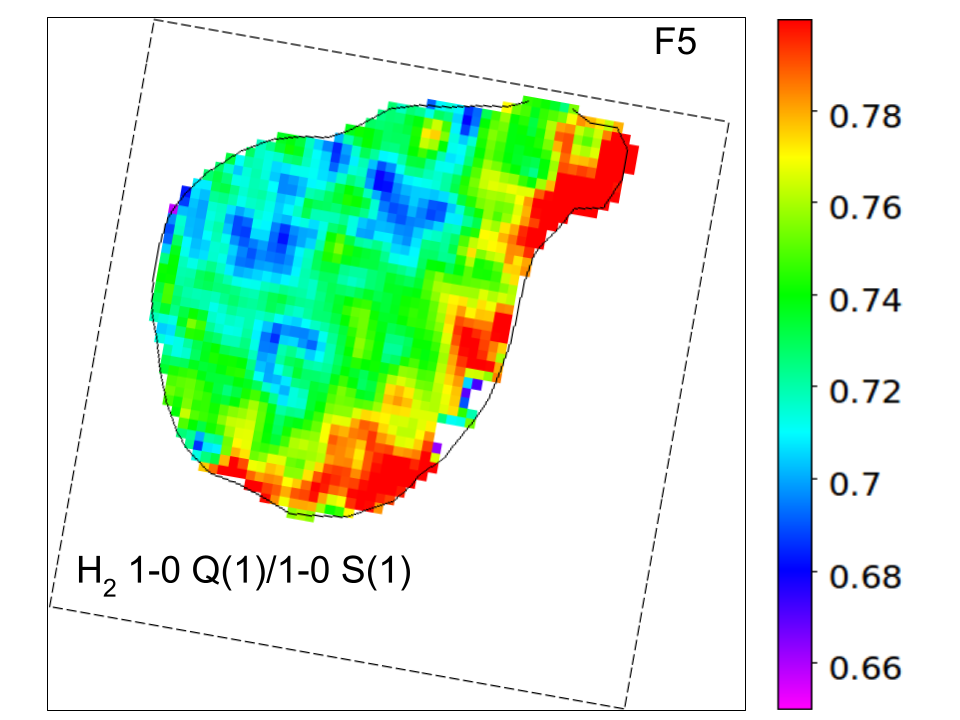}
    \caption{H$_{2}$ 1--0/2--1 S(1), 1--0/S(3) 3--2 S(1) and 1--0 Q(1)/1--0 S(1) ratios in Field\,5. A contour (above 4$\sigma$) of the H$_{2}$ 1--0 S(1) emission (see Fig.\,\ref{mapsF5}) is displayed for reference. }
              \label{ratios5}
    \end{figure}

The richness of H$_{2}$ lines and the good quality of the data in Field\,5 allow us to perform a line ratio study with the 
aim of establish the nature of such molecular emission that presents such curious morphology in the context of a likely end of a jet. Following \citet{martin-h08} and \citet{chen15}, we present H$_{2}$ 1--0/2--1 S(1), 
1--0 S(1)/3--2 S(3), and 1--0 Q(1)/1--0 S(1) ratios to distinguish between collisional and radiative H$_{2}$ excitation (see Fig.\,\ref{ratios5}).
It is important to be cautious with unreal low values
in the H$_{2}$ 2--1 S(1), 3--2 S(3), and 1--0 Q(1) emissions that will yield artificially high values in the corresponding ratios. Thus, only values above 5$\sigma$ for such emissions were used. The average values along the whole field of the presented ratios are 11, 50, and 0.74, respectively, 
and the value ranges (min.--max.) of each ratio are: 6.6--16.5, 17.8--94.8, and 0.65--0.87, respectively. The meaning of these values is discussed in Sect.\,\ref{f5}.

\section{Discussion}
\label{discussS}

As presented in Fig.\,\ref{present}, MYSO G79 presents at near-IR bands a striking morphology with knots and arcs extending southwards the compact object (i.e. the massive protostar). 
Thus, this is an interesting case to study with great detail the physical consequences of such a jet in the MYSO close surroundings. The Gemini NIFS data allow us to perform a kinematical and a detailed qualitatively morphological study of the continuum and emission lines, and from the evaluation
of line ratios we can discern the line excitation mechanisms. In what follows, we discuss the main results in each observed field, and we finish with a comprehensive view of the possible dynamics of the jet trying to explain such a complex morphology.

\subsection{Field 1}
\label{F1}

The protostar, understood as the compact object responsible of the jets, lies at the peak of the 2.2 $\mu$m continuum emission (see Fig.\,\ref{mapsF1} top panel). The continuum emission morphology is elongated towards the south and it separates into two branches. These
features are contained in a region of about 1 arcsec  (1400 au at the distance of 1.4 kpc). It is known that continuum emission at 2.2 $\mu$m around YSOs can be explained as an scattered light nebulosity, in which the light scattering process occurs in the walls of a cavity that was cleared out in the circumstellar material by a jet (e.g. \citealt{bik06}). This is supported by the Br$\gamma$ and H$_{2}$ emissions (see Fig.\,\ref{mapsF1} middle and bottom panels), that present some morphological correspondence with both branches of the continuum emission. One of these branches (the southernmost) would seem to be resolved in a probably point-like source with some extended Br$\gamma$ emission. This possibility is discussed in Sect.\,\ref{jk}.

The hydrogen Br$\gamma$ emission line is commonly observed towards massive YSOs (e.g. \citealt{cooper13}). While the majority of Br$\gamma$ emission detected in 
the spectra of YSOs is believed to arise from the recombination regions associated with the magnetospheric accretion of circumstellar disk material onto the
forming star \citep{beck10,guo21}, an additional emission component originating from the strong winds driven by the massive protostar should be important \citep{davis10}. 
The Br$\gamma$ map that the integral field spectroscopy allows us to construct is useful to study this issue. In our case, the Br$\gamma$ peak very likely traces the 
accretion processes, but the extended emission, with such morphological correspondence with the continuum emission, strongly suggests that in this region the Br$\gamma$ line arises from stellar strong winds \citep{fedriani19}, which probably have cleared out the mentioned cavity.

The H$_{2}$ 1--0 S(1) line emission presents an interesting morphology: two cone-like features extending to the north and south, respectively, with their respective vertices pointing to the continuum peak (i.e. the position of the protostar). It is important to remark that unlike the MHO catalog \citep{makin18} in which it is indicated that this source presents
only one H$_{2}$ lobe whose emission is observed as a collection of knots (the southern emission along the different observed fields in this work), we also detected northwards H$_{2}$ emission.
The excitation of this line can be due to collisions or to UV fluorescence. Ratios among different H$_{2}$ near-IR lines are usually used to discern between such excitation mechanisms (see the case of Field\,5). However, in Field\,1 the 1--0 S(1) is the only H$_{2}$ line that appears, thus, following \citet{hatch05} and \citet{chen15} we evaluate the H$_{2}$ 1--0 S(1)/Br$\gamma$ ratio (see Fig.\,\ref{ratio1}).
We obtain ratios $\leq$1 (with an average of 0.6) along the analyzed region, suggesting a stellar UV excitation mechanism for the H$_{2}$. This is in agreement with the H$_{2}$ double cone-shape morphology and the likely cavities carved out in the circumstellar material: the molecular gas lying at the cavities walls is excited by the UV photons from the protostar. Additionally, it is interesting to note a gradient from the center to the north and southern borders in such a ratio, reaching to values slightly larger than 1. This can be showing two bipolar thin regions where the H$_{2}$ excitation may have a collisional contribution. 

Finally, it is important to note the absence of CO band-heads at 2.3--2.4 $\mu$m nor in emission neither in absorption. Even though these CO features are usually detected
towards YSOs (\citealt{hoff06,ilee13,fedriani20} and references therein) it is not rare their non detection. As \citet{hoff06} mention, the weakness or even the absence of any 
CO features may therefore be a complex combination of various competing mechanisms. In our case, as in some YSOs studied in \citet{martins10}, the source presents the combination of H$_{2}$ 1--0 S(1) and Br$\gamma$ emissions with the absence of CO band-heads features. \citet{hoff06} found among their large sample of analyzed massive YSOs that the CO featureless objects are basically later than B3, and suggest that they may be similar to other hot YSOs like the Herbig Ae/Be star AB Aurigae \citep{hart89}. From observations of Herbig Ae/Be stars, \citet{kraus08} suggested that the Br$\gamma$ emission originates in extended stellar or disk winds, which is in agreement with our interpretation of such an emission.

\subsection{Fields 2, 3, and 4}

Fields\,2,\,3, and\,4 show continuum emission in decreasing intensity (Figs.\,\ref{mapsF2} top, \ref{mapsF3}, and \ref{mapsF4}) and, besides Field\,2 that shows an intense H$_{2}$ 1--0 S(1) bulk of emission (Fig.\,\ref{mapsF2} bottom), these fields do not present emission lines.

In these fields the continuum emission shows arc-like features that can be resolved into some maximums. In Field\,2 the continuum emission presents a zigzagging morphology with three maximums, from which two of them coincides with H$_{2}$ 1--0 S(1) 
emission (see Fig.\,\ref{mapsF2}). Given that this is the only H$_{2}$ line that appears in the region we cannot study its excitation mechanism. However, based on the observed knots morphology, that usually it is found in the shocked gas in HH objects 
(e.g. \citealt{davis94,coey04}), we suggest that this emission has a collisional origin due to the passage of a jet which generated a cavity observed in the continuum emission. The continuum zigzagging morphology can be considered as a superposition of two arc-like features, with their concave curvatures pointing to the north-west and east, respectively, whose radii size is about 
1200 au in both. Strikingly these features resemble to the cork-screw like structures obtained in computational studies of disk winds, jets,
and outflows \citep{pudritz07,staff15}. We wonder if we are observing at the near-IR bands the consequences of a jet with a complex kinematics. 
This will be discussed below.

In Fields\,3\,and\,4 the near-IR continuum emission also presents an arc-like morphology. In Field\,3, an arc with the concave curvature pointing to the north appears in perfect correspondence with the UKIDSS data (see Figs.\,\ref{mapsF3} and\,\ref{obs}). Our Gemini data allow us to resolve this arc-like feature into two branches. 
The radius size of this arc is about 2000 au. In the case of Field\,4, while the near-IR continuum emission also has a morphology of arc with its 
concave curvature pointing to the north, it also seems like a filament. These structures reinforces the hypothesis of the presence of a cork-screw like feature due to the action of a jet.

\subsection{Field 5}
\label{f5}

Field\,5 is characterized by weak continuum emission and the richness of H$_{2}$ lines (see spectrum in Fig.\,\ref{spectF5}). The maps (see Figs.\,\ref{mapsF5} and \ref{mapsF5cont}) show that the lines emission does not correlate with the continuum, and they present a circular clump structure. Given that this field could be the end of the jet, we wonder if such a structure is due to shocked molecular gas in a bow shock feature produced by the jet (e.g. \citealt{oconnell04}). The analysis of the line ratios presented in Sect.\,\ref{ratioSect} strongly suggests the collisional origin of the H$_{2}$ emission. 
As explained in \citet{martin-h08}, our average values of 11 and 50 obtained in the H$_{2}$ 1--0/2--1 S(1) and 1--0 S(1)/3--2 S(3) ratios, respectively, indicate shocked gas. This is reinforced by the average value of 0.74 obtained in the H$_{2}$ 1--0 Q(1)/1--0 S(1) ratio. Shocked thermal models with a temperature of 2000 K predict a value of 0.7 in this ratio \citep{luhman98,chen15}. Thus, given the observed morphology of the molecular emission in this field, 
it is very likely that we are observing some part of the surface of a bow shock feature projected in the plane of the sky. 
According to \citet{gust10}, the morphology of a bow shock projected onto the plane of the sky naturally depends on the viewing angle and on 
the orientation of the magnetic field. The authors also show that the line brightness and line ratios can change with viewing
angle. In the following section (Sect.\,\ref{jk}) we discuss the jet morphology and geometry.

The radial velocity measured for this very likely bow shock is v$_{\rm LSR} \sim -31$ \ks. Taking into account that 
the systemic velocity of the source is  v$_{\rm LSR} = -1.7$ \ks~(\citealt{urqu11}; and see Sect.\,\ref{large}), we conclude that
we are observing shocked gas coming to us. Thus, our Gemini data is tracing a blue-shifted jet with a velocity of about $28/cos(i)$~\ks, where $i$ is the inclination angle of the jet respect to the line of sight.

Assuming that indeed we are observing a bow shock in Field\,5, or some part of it, by analyzing the moment 1 map presented in Fig.\,\ref{m1H2F5}, in which
the lowest velocities (i.e. more negative) are observed almost at the center of the structure (i.e. probably the apex of the bow feature), we conclude that the jet is indeed coming to us with some inclination along the line of sight. Moreover, the moment 1 map shows a conspicuous velocity gradient with radial velocities increasing from the center towards the border of the structure. Interestingly, towards the north rim it can be noticed a smaller gradient (green border), which would indicate that the bow shock apex would be pointing towards the south. This radial velocity behaviour of the bow shock related to MYSO G79 is in agreement with the results exhibited in Figure 10 of the work of \citet{gust10}.

It is interesting to note that in most of the H$_{2}$ maps, the circular clump feature present a clumpy substructure with two maximums, one to the north, and the other one, the most intense, almost lying at the center. Structured bow shocks are usually observed in HH objects
\citep{hart11}, and such structures that appear as clumps along the bow shock could be due to several phenomena: thermal instability triggered by strong radiative cooling in the shock, Kelvin–Helmholtz instability, or even a clumpy preshock density \citep{sukui15,hansen17}. 

Even though the ratios presented in Sect.\,\ref{ratioSect} are quite uniform across the emitting area indicating a uniform mechanism for the H$_{2}$ emission, they present some variation along the observed structure. In no case do these variations change the interpretation of the H$_{2}$ excitation mechanism, but they may indicate some different physical conditions in the molecular gas that belongs to the bow shock feature, or they could be consequence of the bow shock viewing angle \citep{gust10}.  Precisely, by inspecting the results presented by these authors from their 3D model of bow shocks,  we find some morphological correspondences with our results. In particular, the modeled morphologies of the H$_{2}$ 1--0 S(1) emission, 1--0/2--1 S(1) ratio and the radial velocity distribution of a bow shock observed with an inclination angle along the line of sight, $i$, between 20\d~and 50\d, with $\theta$, an angle related to the direction of the assumed uniform magnetic field, between 0\d~and 30\d, are similar to our observational results. Taking into account that \citet{gust10} focused in predicting line emission maps of molecular hydrogen in C-type bow shocks, this comparison is in agreement with the fact that we are observing a C-type shock from a YSO jet with a velocity of about $28/cos(i)$~\ks~($30-43$ \ks, by considering the $i$ angle range mentioned above), and hence, we are presenting observational evidence that can support the models presented by these authors.

\subsection{Jet morphology, kinematics, and possible causes}
\label{jk}

\begin{figure}[h]
   \centering
   \includegraphics[width=8cm]{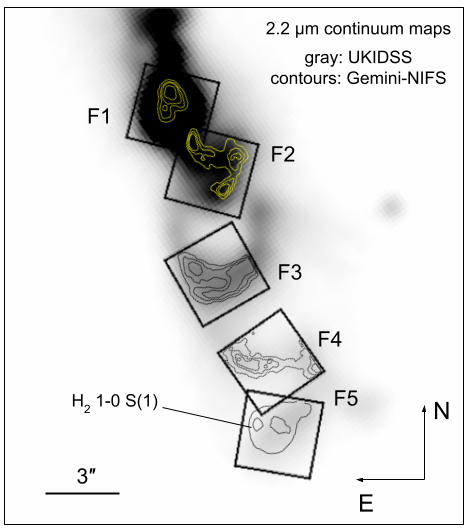}
    \caption{In gray it is presented Ks-band emission obtained from UKIDSS, and the contours in each field are the continuum at 2.2 $\mu$m obtained from the NIFS data, except in Field\,5 where some contours of the H$_{2}$ S(1) 1--0 emission are displayed.  For a better display, the emissions observed with NIFS were slightly smoothed with a boxcar function with a factor of 2.} 
              \label{general}
    \end{figure}

In this section we discuss the morphology of the observed near-IR emission and what it can inform us about the kinematics
of the jet. Fig.\,\ref{general} displays the UKIDSS Ks emission with contours of the Gemini NIFS data to appreciate a general
view of the studied near-IR structure. The continuum at 2.2 $\mu$m obtained with NIFS resolves some features observed
in the UKIDSS data, and the general view allow us to suggest that the jet is precessing. The observed features
at the near-IR continuum resemble to the cork-screw like structures obtained and observed in both computational and observational studies
of likely precessing jets (e.g. \citealt{rosen04,mov07,paron16,beltran16,ferrero22}).
For instance, \citet{smith05} pointed out that the dominant structure produced by a precessing jet is an inward-facing cone, and particularly, a slow-precessing jet leads to helical flows, generating a spiral shaped nebula as observed in our case.
This kind of structures are also obtained in simulations of magnetohidrodynamic disk winds \citep{pudritz07,staff15}. Indeed, the role of magnetic fields can be important in
the dynamics of the disk and jets \citep{pudritz19}, however, if one proves a tidal interaction in a binary system, gravity should be 
the dominant mechanism explaining the jet precession. 

We measure the projected wavelength ($\lambda_{p}$) of the helical pattern \citep{terquem99} by considering the distance between the first curved feature in Field\,1 and that of Field\,2, obtaining 7000 au. An estimate of the actual wavelength is obtained from $\lambda = \lambda_{p}/{\rm sin}(i)$. Considering that the jet velocity can be estimated from $v_{jet} = \lambda/\tau_{p}$, using $v_{jet}$ between 30 and 43 \ks~(see the end of Sect.\,\ref{f5}), we obtain a $\tau_{p}$ about $(1.1-1.8)\times$10$^{3}$ yr, which is in agreement with a slow-precessing jet \citep{smith05}.

The observed angular length of the jet at near-IR, measured from the center of Field\,1 to the bow shock observed in Field\,5, is about 16 arcsec, which gives a projected length of 0.1 pc at the distance of 1.4 kpc. By considering an inclination angle between 20\d~and 50\d~respect to the line of sight (see Sect.\,\ref{f5}), we conclude that the length of the jet would be 0.13--0.30 pc, which is an interval of usual values measured in YSO jets \citep{samal18}. 

\begin{figure}[h]
   \centering
   \includegraphics[width=7cm]{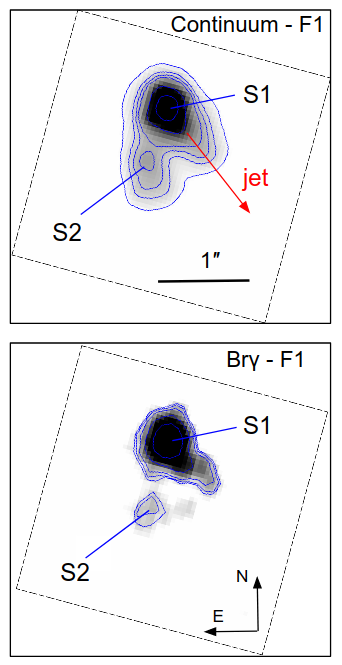}
    \caption{Continuum and Br$\gamma$ emission towards Field\,1 displayed in a gray scale with some contours that remark the presence of two peaks, likely two components of a binary system. Such components are called S1 and S2, and the direction of the jet, emerging from S1, is indicated with a red arrow.  The separation between S1 and S2 is 0\farcs7~(a projected distance of about 980 au at the assumed distance of 1.4 kpc). For a better display, both emissions were slightly smoothed with a boxcar function with a factor of 2. The first contours of the continuum and Br$\gamma$ emissions are 24$\sigma$ and 4$\sigma$, respectively. }  
              \label{bin}
    \end{figure}

By carefully inspecting the 2.2 $\mu$m continuum and Br$\gamma$ emission in Field\,1 (see Fig.\,\ref{bin}), in which another 
peak, at both emissions, appears over one of the described branches in Sect.\,\ref{F1}, we suggest that 
MYSO G79 may be composed by a binary system (probably S1 and S2 in the figure). Both sources are separated by 0\farcs7~(a projected distance of about 980 au at the assumed distance to MYSO G79 of 1.4 kpc). Binarity seems to be something common in MYSOs, and the observed projected binary separation in this case (a lower limit of the actual
separation) is within the physical separation range observed in many cases \citep{pomo19}.
Thus, we suggest that the extended near-IR features should be due to a precessing jet that can be explained through the tidal interaction between the companion stars. However, we cannot discard that the secondary peak of the Br$\gamma$ emission could be due to a jet knot from the main source, and in this case, if the cause of the jet precession is a binary system, the companion stellar object may be hidden or not resolved in our image.

Radio continuum emission can shed light about the nature of the central stellar object and can give us limits for its spectral type. \citet{srid2002} detected radio continuum emission at 3.6~cm towards MYSO G79 (S$_{3.6}\sim$1.4~mJy). If it assumed that this emission arises from ionized gas of an optically thin hyper- or ultra-compact \hii~region associated with a young massive star, based on the estimated radio flux density, we can conjecture upon the spectral type
of the young star that is generating the observed jet. The number of photons needed to keep a compact  \hii~region ionized, in an optically thin regime, is given by N$_{\rm uv} = 0.76 \times 10^{47}~{\rm T_4^{-0.45}~
{\nu_{GHz}}^{0.1}~S_{\nu}~D_{kpc}^2}$ \citep{chai1976}, where T$_4$
is the electron temperature in units of $10^4$~K, D$_{\rm kpc}$ is the distance in kilo-parsecs, $\nu_{\rm GHz}$ is the frequency in GHz, and S$_\nu$ is the measured total flux density in Jansky. We assumed an electron temperature of T=10$^4$~K and a distance of 1.4~kpc. 
We derived a total amount of ionized photons of about
${\rm N_{uv} = 2.6 \times 10^{46}~ph~s^{-1}}$. Based on \citet{ave1979} and
\citet{mar2005}, the spectral type of the possible young exciting star is later than a B0V.  In the case of a binary system, this estimation may 
correspond to the main source (i.e. the most massive star) of the system, or indicates that the ionization generated by both stars is an analogous to a star later than B0V.

On the other hand, from more recent observations with higher angular resolution, \citet{rosero2019} based on the low radio luminosity ($S_{\rm 5~GHz} d^2 \sim 0.15~{\mu}{\rm Jy~kpc}^2$), the spectral index ($\sim$ 0.9), the relatively elongated morphology,  and its alignment with the direction of the H$_2$-jet, suggested that the radio continuum emission associated with MYSO G79 arises from an ionized jet.

For a complete understanding of the spatial configuration of any jet, and hence of the associated extended molecular outflows, it is necessary to be sure what is the component coming to us (blue-shifted) and which one is moving away from us (red-shifted), naturally, with some inclination along the line of sight. \citet{maud15b}, who studied outflows in a
large sample of sources, based on CO emission showed that MYSO G79 has a red-shifted molecular outflow extending towards the southwest (see their Figure\,C1). Thus, based on our findings, we decide to re-analyze such CO data, which is presented and discussed in the following section.

\subsection{Molecular gas at a larger spatial scale}
\label{large}

In this section we discuss which was obtained by re-analyzing millimeter data from the JCMT and IRAM telescopes with the aim to probe the molecular gas in which MYSO G79 is embedded and likely is affecting. 

\begin{figure}[h]
   \centering
   \includegraphics[width=9cm]{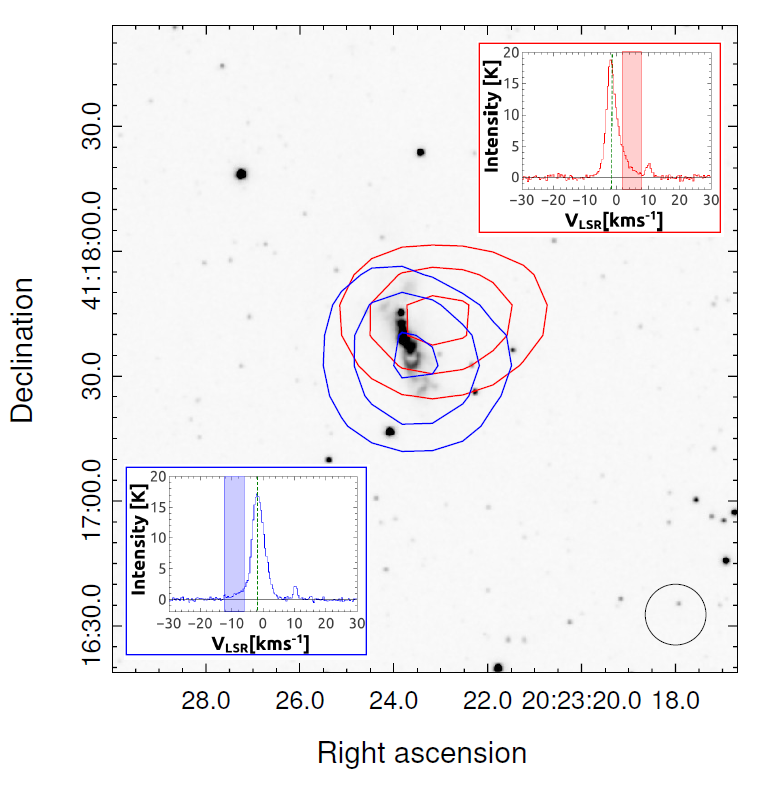}
    \caption{UKIDSS image at Ks band of MYSO G79. The blue and red
contours represent the JCMT $^{12}$CO J = 3--2 emission averaged from -6 to -12
kms$^{-1}$ (blue lobe), and from 2 to 8 km s$^{-1}$ (red lobe), respectively. The
blue contours are at 1, 2, and 3~K and the red ones are at 2, 3, and 4~K. The $\sigma$ noise level is 0.3 K.  The averaged spectra (in a beam area) towards the center of both features are shown. The vertical dashed green lines represent the systemic velocity of the source (at about $-1.7$ \ks). The circle of 14\farcs5~in diameter at the bottom right corner represents the JCMT beam.}
              \label{outflows}
    \end{figure}

We re-analyze the CO data obtained from the JCMT telescope that were presented in \citet{maud15b} in order to investigate in detail molecular outflows related to MYSO G79 (for the description of such data, see \citealt{maud15b}). We found that the spectra of the \2 data lying towards the north west of MYSO G79 present a red-shifted wing at the velocity range $+2$ to $+8$ \ks, while the \2 spectra that lie towards the south of G79 has a blue-shifted wing at the velocity range $-6$ to $-12$ \ks.
Figure\,\ref{outflows} displays two spectra showing such spectral wings and the averaged CO emission in these velocity ranges is presented in blue and red contours over the UKIDSS near-IR image. The peak of the blue-shifted CO feature coincides with the southern near-IR structure, confirming that the jet is coming to us along the line of sight. Our finding, regarding to the location of the blue- and red-shifted molecular gas, is the opposite of what was presented in \citet{maud15b}, and the reason must be that the authors, in the case of this source, included some bulk molecular gas in the determination of the outflows.

\begin{figure*}[h]
   \centering
   \includegraphics[width=7cm]{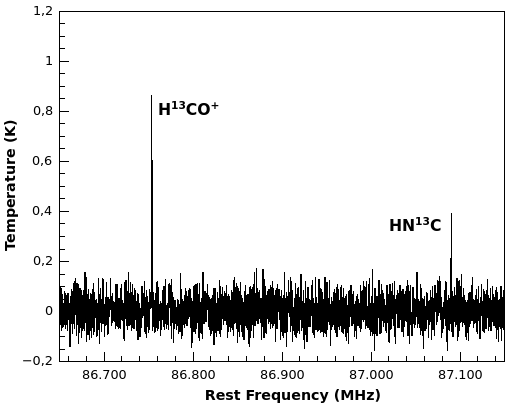}
  \includegraphics[width=7cm]{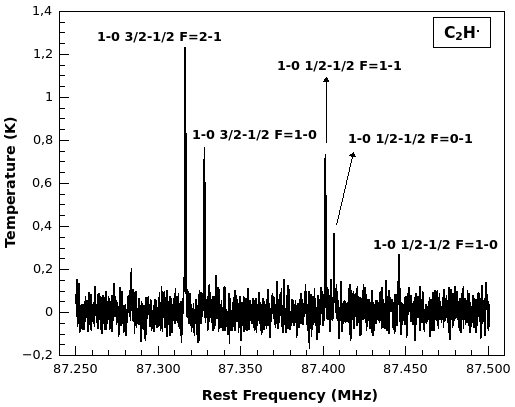}
  \includegraphics[width=7cm]{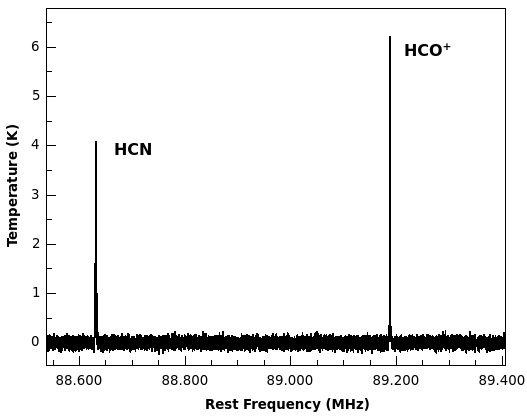}
 \includegraphics[width=7cm]{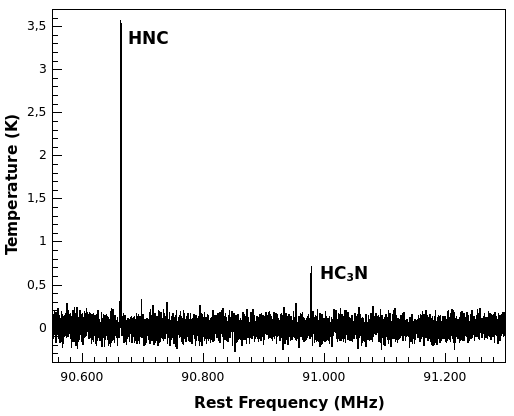}
   \includegraphics[width=7cm]{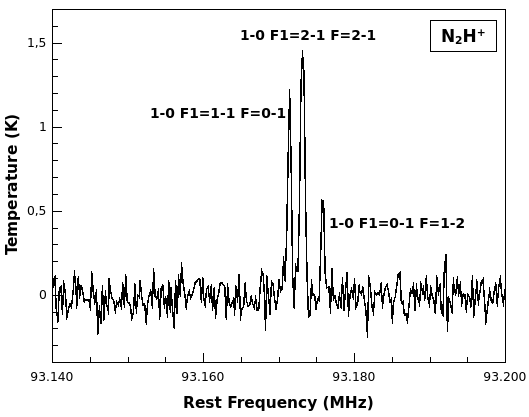}
   \includegraphics[width=7cm]{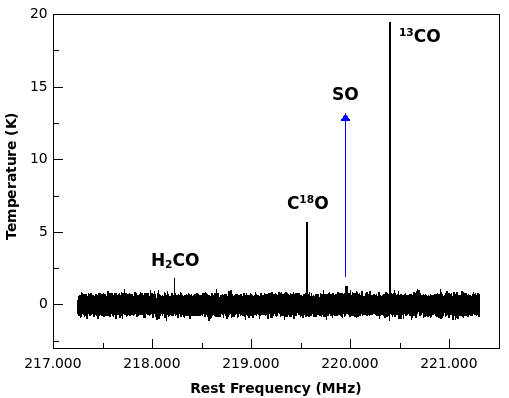}
      \caption{Spectra obtained from the IRAM 30\,m Telescope \citep{gerner14} towards the region where MYSO G79 is embedded. The identified molecular species are indicated in each spectrum. In case
      of C$_{2}$H and N$_{2}$H$^{+}$ the hyperfine lines are also indicated.}
              \label{iram}
    \end{figure*}

\citet{gerner14}, using data from the IRAM 30\,m Telescope, studied the chemistry of 59 high-mass star forming regions, sample in which MYSO G79 is included.
Thus, using these data retrieved from the VizieR Catalog (catalog: J/A+A/563/A97), we analyze with some more detail the molecular gas, and its chemistry, of the region in which G79 is embedded.
The IRAM data consist in an observation of four single spectra at the frequency ranges: 86--90, 90--94, 217--221, and 241--245 GHz centered at RA$=$20:23:23.8, dec.$=$+41:17:40.0 (J2000) (almost the center of Field\,1). In the case of this source, none molecular line appears in the last frequency range. The beam of the observations at the first two frequency ranges is about 29\s, while at the others frequency ranges is about 11\s. Thus,
in the first case the spectra probe gas of a region that completely contains MYSO G79 and its jets, while the other spectra probe gas mainly towards the center of the region where the protostar lies (Fields 1 and 2, and the same extension towards the north). 

Using the molecular rest frequency database from the National Institute of Standard and Technology (NIST)\footnote{https://physics.nist.gov/cgi-bin/micro/table5/start.pl}, we successfully identified all molecular emission lines that appear in the spectra (see Fig.\,\ref{iram}), coinciding with those presented by \citet{gerner14} in relation to the source named HMPO20216. Besides that we remark the hyperfine lines of C$_2$H and N$_2$H$^{+}$, as a novelty, we report the presence of the cyanoacetylene (HC$_3$N) at 91.97 GHz, which was not taking into account by \citet{gerner14}. 
Given the chemical richness that stands in MYSO G79, we use some molecular lines to obtain information in order to analyze whether the activity of the
jets and outflows has influenced the chemistry in the region. 

The comparison between lines velocity widths of some molecular species is useful to analyze the evolutive stage of the region (e.g. \citealt{yu15}). From this set of data, from gaussian fittings we measured the FWHM $\Delta$v$^{\rm H^{13}CO^{+}}$,  $\Delta$v$^{\rm C_{2}H}$, $\Delta$v$^{\rm HCO^{+}}$, $\Delta$v$^{\rm HC_{3}N}$, 
and $\Delta$v$^{\rm N_{2}H^{+}}$ in 1.36, 1.75, 2.35, 1.33, and 2.11 \ks, respectively. As it was found by \citet{yu15}, the velocity widths of HC$_{3}$N and C$_{2}$H are similar to that of H$^{13}$CO$^{+}$, and the obtained ratios $\Delta$v$^{\rm C_{2}H}$/$\Delta$v$^{\rm HC_{3}N} = 1.31$ and  $\Delta$v$^{\rm N_{2}H^{+}}$/$\Delta$v$^{\rm HC_{3}N} = 1.58$ indicate a MYSO in which the effects of the inner jets or the radiation of a possible incipient \hii~region are not still evident. As the authors propose, it seems that N$_{2}$H$^{+}$ and C$_{2}$H emissions do not come from the stirred-up gas in the center of the clump.

Additionally, from the isomers HCN and HNC (cyanide and isocyanide hydrogen, respectively) we can estimate the gas kinetic temperature (T$_{\rm k}$) over the studied region. From the integrated intensity ratio (I(HCN)/I(HNC)), which in our case is about 2, it is possible to estimate T$_{\rm k}$ using the expression proposed by \citet{hacar20}. Such an empirical correlation proposed by the authors yields T$_{\rm k} \sim 20$ K for the region in which MYSO G79 is embedded, indicating that these molecular lines are probing cold gas, likely from the envelope in which the source is embedded. 
If we assume that the I(HCN)/I(HNC) ratio is proportional to the abundance ratio, from the obtained value, and following \citet{hacar20} we can say something about the chemical reactions governing these molecular species. We point out that slow HNC destruction is occurring at this region with T$_{\rm k}$ about 20 K, and hence, the predominant mechanism seems to be the neutral-neutral reaction of this molecule with oxygen (HNC + O $\rightarrow$ NH + CO) rather than involving H mechanism which is relevant at T$_{\rm k}$ $>$ 40 k (HNC + H $\rightarrow$ HCN + H). The energy barrier of the former reaction has been studied by \citet{hacar20} and it is proposed to be $\Delta E \sim 20$ K.  

We conclude that the complex chemistry revealed by the detection of these molecular species is not a consequence of the jet activity. We point out that these molecular lines probe gas of the external layers of the clump in which MYSO G79 is embedded. This is indeed a very interesting source to be mapped with very high-angular resolution at (sub)millimeter
wavelengths, for instance, using the Atacama Large Millimeter Array (ALMA).

\section{Summary and concluding remarks} 
\label{concl}

Using near-IR integral field spectroscopy, radio continuum, and millimeter data we studied in detail MYSO G79.
The analysis of a jet extending southwards the source shows cork-screw like structures at 2.2 $\mu$m continuum, strongly suggesting that the jet is precessing. The obtained velocity of such a jet is between 30 and 43 \ks~and it is blue-shifted, i.e. the jet is coming to us along the line of sight. We suggest that the precession may be produced by the gravitational tidal effects generated in a probable binary system, and we estimate a jet precession period of about 10$^{3}$ yr, in agreement with a slow-precessing jet, which can explain the observed helical features.
We presented a detailed map of a bow shock produced by such a jet observed coming to us with some inclination along the line of sight. Given that in the literature there are a few works reporting bow shocks from jets driven by MYSOs (e.g. \citealt{fedriani18}), we point out that we are presenting an interesting source in which a bow shock generated by a MYSO jet is observed (and resolved) almost from the front.  We conclude that we are presenting an interesting observational piece of evidence that can support theoretical models of bow-shocks and precessing jets, or can be used to contrast or to probe such models.

Additionally, we found that the molecular outflow generated by the investigated jet is coming to us, and it is not going away from us as was stated in a previous work \citep{maud15b}. A brief analysis of several molecular species indicates a complex chemistry developing at the external layers of the molecular clump in which G79 is embedded, and it seems that the jet activity have not any influence in such a chemistry yet. This is a very interesting source to be mapped with ALMA with the aim to resolve the molecular outflows and to study the chemistry of the deepest gas in the clump related to the MYSO activity. 

We conclude that in addition to the large MYSOs and outflows surveys, dedicated studies of
particular sources like this, in which the observations (mainly at near-IR and
(sub-)mm) are analyzed in deep, are extremely useful to shed
light on the processes involved in the formation of massive stars and their consequences in the surrounding interstellar medium.

\begin{acknowledgements}

We thank the anonymous referee for her/his useful comments and suggestions. Based on observations obtained at the international Gemini Observatory, a program of NSF’s NOIRLab, which is managed by the Association of Universities for Research in Astronomy (AURA) under a cooperative agreement with the National Science Foundation. on behalf of the Gemini Observatory partnership: the National Science Foundation (United States), National Research Council (Canada), Agencia Nacional de Investigaci\'{o}n y Desarrollo (Chile), Ministerio de Ciencia, Tecnolog\'{i}a e Innovaci\'{o}n (Argentina), Minist\'{e}rio da Ci\^{e}ncia, Tecnologia, Inova\c{c}\~{o}es e Comunica\c{c}\~{o}es (Brazil), and Korea Astronomy and Space Science Institute (Republic of Korea). This work was partially supported by grants PICT 2015-1759 and PICT 2017-3301 awarded by Foncyt, and grant
PIP 2021 11220200100012 by CONICET. M.B.A. and N.C.M. are doctoral fellows of CONICET, Argentina. S.P., D.M., and  M.O. are members of the Carrera del Investigador Cient\'\i fico of CONICET, Argentina. 

\end{acknowledgements}


%%%%%%%%%%%%%%%%%%%%%%%%%%%%%%%%%%%%%%%%%%%%%%%%%%%%%%%%%%%%%%%%%%%%%
\bibliographystyle{aa}  % A&A format
   %\bibliographystyle{klunamed}     
   % format of references provided by the review (.bst)
\bibliography{ref}
   % file containing the bibtex references (.bib)
\IfFileExists{\jobname.bbl}{}
{\typeout{}
\typeout{****************************************************}
\typeout{****************************************************}
\typeout{** Please run "bibtex \jobname" to optain}
\typeout{** the bibliography and then re-run LaTeX}
\typeout{** twice to fix the references!}
\typeout{****************************************************}
\typeout{****************************************************}
\typeout{}
}
\label{lastpage}

\end{document}